\PassOptionsToPackage{table,xcdraw}{xcolor}

\documentclass[acmlarge,screen]{acmart}

\AtBeginDocument{%
  \providecommand\BibTeX{{%
    \normalfont B\kern-0.5em{\scshape i\kern-0.25em b}\kern-0.8em\TeX}}}
    
\usepackage{longtable}
\usepackage{breakcites}
\usepackage{tcolorbox}
\usepackage{enumitem}
\usepackage{mathtools}
\newcommand{\myitem}[1]{\vspace{0.25\baselineskip}\noindent\textbf{#1}}
\newcommand{\tabitem}{~~\llap{\textbullet}~~}
\usepackage[normalem]{ulem}
\useunder{\uline}{\ul}{}
\usepackage{multirow}
\usepackage{tabularx}
\newcolumntype{a}{>{\columncolor[HTML]{f2f2fb}}l}
\newcolumntype{d}{>{\columncolor[HTML]{f2f2fb}}c}
\usepackage{pbox, cellspace}
\newcommand{\rev}[1]{\textcolor[rgb]{0.00,0.00,0.00}{#1}}

\AtBeginDocument{%
  \providecommand\BibTeX{{%
    \normalfont B\kern-0.5em{\scshape i\kern-0.25em b}\kern-0.8em\TeX}}}

\setcopyright{acmcopyright}
\copyrightyear{2021}
\acmYear{2021}
\acmDOI{10.1145/1122445.1122456}

\acmBooktitle{ACM Transactions on Computing for Healthcare}

\begin{document}

\title[14 Years of Self-Tracking Technology for mHealth - Literature Review: Lessons Learnt and the PAST SELF Framework]{14 Years of Self-Tracking Technology for mHealth - Literature Review: Lessons Learnt and the PAST SELF Framework}
\author{Sofia Yfantidou}
\email{syfantid@csd.auth.gr}
\author{Pavlos Sermpezis}
\email{sermpezis@csd.auth.gr}
\author{Athena Vakali}
\email{avakali@csd.auth.gr}
\affiliation{%
  \institution{Aristotle University of Thessaloniki}
  \city{Thessaloniki}
  \country{Greece}}


\begin{abstract}
In today's connected society, many people rely on mHealth and self-tracking (ST) technology to help them adopt healthier habits with a focus on breaking their sedentary lifestyle and staying fit. However, there is scarce evidence of such technological interventions' effectiveness, and there are no standardized methods to evaluate their impact on people's physical activity (PA) and health. This work aims to help ST practitioners and researchers by empowering them with systematic guidelines and a framework for designing and evaluating technological interventions to facilitate health behavior change (HBC) and user engagement (UE), focusing on increasing PA and decreasing sedentariness. To this end, we conduct a literature review of 129 papers between 2008 and 2022, which identifies the core ST HCI design methods and their efficacy, as well as the most comprehensive list to date of UE evaluation metrics for ST. Based on the review's findings, we propose PAST SELF, a framework to guide the design and evaluation of ST technology that has potential applications in industrial and scientific settings. Finally, to facilitate researchers and practitioners, we complement this paper with an open corpus and an online, adaptive exploration tool for the PAST SELF data.
\end{abstract}



\maketitle

\section{Introduction}\label{introduction}
Nowadays, technology is an inextricable part of our lives that has affected us in multiple ways. 
Naturally, the field of health and well-being has not been left untouched. Especially when viewed through the lens of health promotion, technology is like a double-edged sword. On the one hand, research shows that technology may be detrimental to people's mental health and well-being \cite{cramer2018statusofmind}, while it has also contributed to a decline in physical activity (PA) by promoting a more sedentary lifestyle \cite{nigg2003technology}. On the other hand, technological advancements and mHealth have enabled individualized health-promoting interventions to large populations via differing channels \cite{nigg2003technology}. For instance, electronic health records, remote monitoring, and digital diagnostics are revolutionizing the healthcare domain by enabling more integrated, effective, and faster care even from a distance. 

\myitem{\rev{mHealth: the role of PA.}} Simultaneously, technological innovations aim to tackle the root causes of the population's health problems. 
\rev{According to the World Health Organization (WHO), physical activity is amongst the determinants of good health\footnote{``Determinants of health.'' 3 Feb. 2017, \url{https://www.who.int/news-room/q-a-detail/determinants-of-health}. Accessed 28 Sept. 2021.}.
Specifically, physical inactivity has been identified by the WHO as the fourth leading risk factor for global mortality, accounting for 6\% of deaths globally \cite{world2010global}. At the same time, according to \citet{Epstein2020}, it is the most well-studied domain within mHealth, accounting for more than 1 out of 3 publications in the related literature. Hence, it is evident that the concept of mHealth is inextricably linked to PA and PA-promoting technologies.} At the individual level, physically active people enjoy various health benefits, such as improved muscular and cardio-respiratory fitness and lower coronary heart disease rates. At a collective level, more active societies can generate additional returns on environmental and social benefits, such as reduced use of fossil fuels, cleaner air, and healthier economies \cite{world2019global}. Nevertheless, in today's society, fewer and fewer people are sufficiently active, with more than a quarter of the global population not meeting the WHO recommendations for PA\footnote{ "Prevalence of insufficient physical activity - World Health Organization." \url{https://www.who.int/gho/ncd/risk_factors/physical_activity_text/en/}. Accessed 28 Jul. 2020.}. To combat this physical inactivity pandemic, international agencies and organizations are launching short-term campaigns, e.g., United Nations' (UN) PA challenge during the COVID-19 pandemic, or setting long-term goals, e.g., the WHO's Global Action Plan for Physical Activity 2018-2030 \cite{world2019global}. Similarly, technology is trying to address the challenges of this physical inactivity pandemic in multiple ways. For example, social media have provided health experts and aficionados with a platform to share health and PA-related content to billions of users, interactive games, such as Pok\' {e}mon GO, have managed to lift video game players off their sofas \cite{althoff2016influence}, and activity trackers keep us aware of our PA levels at all times. \textit{At the end of the day, though, self-tracking (ST) technology might be the game-changer for health promotion and PA.} 

\myitem{\rev{mHealth: the role of ST.}} \rev{ST (also referred to as self-monitoring, life-logging, quantified self, and personal informatics) refers to ``the practice of gathering data about oneself on a regular basis and then recording and analysing the data to produce statistics and other data (such as images) relating to regular habits, behaviours and feelings'' \cite[\S1]{Lupton2014}. In the digital world, ST refers to the use of ubiquitous technology (such as wearable or mobile devices and apps) for helping users monitor and manage various aspects of their lives, for instance, PA, sleep, and disease.} Unlike desktop computing, ubiquitous computing can occur with any device, at any time, in any place and any data format across any network. Thus, connected wearable devices are becoming the default medium at which health and ST apps are booming\footnote{ "Global connected wearable devices 2016-2022 | Statista." \url{https://www.statista.com/statistics/487291/global-connected-wearable-devices/}. Accessed 28 Jul. 2020.}. As a result, there is a shift from accusing technology of promoting a culture of sedentariness to recognizing mHealth's potential for empowering well-being. 

To accomplish \rev{the mission of assisting users with improving their health outcomes focusing on PA}, ST technology needs to understand the human aspects of interaction with computers. In other words, ST technology needs to achieve two goals: (i) change users' behavior towards healthier habits, referred to as Health Behavior Change (HBC); and  (ii) \rev{monitor these positive changes over time through sustained User Engagement (UE)}. To assist in the fight against physical inactivity, ST technology needs to enable people to successfully change their habits and attitudes to adopt better health behaviors (e.g., increase step counts, decrease prolonged sedentariness). While short-term PA changes can be temporarily beneficial, an efficient ST technology should aim for a long-term HBC. \rev{However, assisting users in their HBC journey requires ways to reliably and accurately measure human behavior and technology usage. Microsoft co-founder and humanitarian Bill Gates (2013) \cite{gates2013bill}, expresses that what is often missing is good measurement and a commitment to follow the data: ``I have been struck again and again by how important measurement is to improving the human condition''. Indeed, designers and researchers need to be able to measure the effectiveness of their interventions to harvest and maintain the full benefits of digital HBC, through successful HCI and UE.} 

\subsection{Motivation: Challenges, Limitations and Technology Status} 
Despite its wide adoption, it is not yet evident whether the current ST technology has succeeded in fulfilling the above requirements. \textit{Do the existing ST solutions achieve lead to successful HBC and UE? Are there more effective techniques than others, and are ST designers aware of them?} The following constraints highlight the current challenges and limitations that motivate our study.

\myitem{Dubious effectiveness of digital HBC.} While ST technology vendors present self-monitoring as the panacea to HBC, there is scattered evidence of the effectiveness of ST technology interventions increasing PA or decreasing sedentariness. Previous literature reviews and meta-analyses reported varied results in improvements in users' PA levels \cite{romeo2019can,fanning2012increasing} , with systems failing many times to reach their full potential \cite{monteiro2019personalization}. Hence, it is unclear how effective ST technology currently is in achieving HBC. According to \citet{aldenaini2020trends}, ST design currently suffers from various limitations: (i) It is unclear which design components lead to increased HBC, i.e., there is no reasoning behind the choice and effectiveness of different system components. (ii) The majority of digital interventions are not theory-grounded and lack a systematic formulation; namely, they do not use any HBC frameworks in the system design, which may limit their effectiveness. (iii) There is no standard approach for evaluating the effectiveness of digital interventions to provide reliable data that can be used for future ST design.

\myitem{High and variable attrition rates.} The health and fitness tech market may be booming, but attrition rates vary from case to case, ranging from 30\% \cite{attrition1} to more than 70\% \cite{attrition2} in some reports. Thus, ST technology may have limited success in actually sustaining UE. \rev{While attrition of a single ST system does not equal abandonment of the desired behavior \cite{rooksby2014personal,Epstein2015b}, this issue highlights the vital need for measuring and evaluating HCI within ST systems, as well as how different technological features and design techniques affect HBC and UE and to what extent. Quantifying UE with ST technology can help us monitor and better assist the users with their HBC journey. In the end, how successful is HCI in ST technology, and how can we measure it?}

Since the design of effective and sustainable ST is a multidisciplinary and complex issue, incorporating theories from Behavioral Science, Computer Science and HCI, and Sports Science, there is a need for specifying the core demands of ST technology. Thus, a clear proposition to encompass the requirements and limitations discussed above should respond to the following critical questions:
\begin{enumerate}[leftmargin=*]
    \item {\textit{Design:}} How can we design effective, evidence-based interfaces and system features for ST technology to encourage HBC and UE?
    \item {\textit{Evaluation:}} How can we standardize the evaluation of the effectiveness of such technologies' interfaces and features to quantify HCI in terms of HBC and UE?
\end{enumerate}

\subsection{Goals and Contributions} 
The two questions above, which we will dismantle into our research questions, have inspired this systematic review. Our vision is to contribute to the understanding of ST technology \rev{for PA} by a systematic review addressing the field's open issues and limitations. To the best of the authors' knowledge, although an abundance of publications refers to the effectiveness of ST interventions for HBC, a systematic analysis that connects the "pieces" between theory and HCI design and evaluation is still missing. The current article is a comprehensive review of 129 studies that examines how ST technology interventions are conducted, designed, and evaluated, covering a period of 14 years (2008–2022) and a total of more than 6 million users across all investigated articles.

\begin{figure}[t]
  \centering
  \includegraphics[width=.8\linewidth]{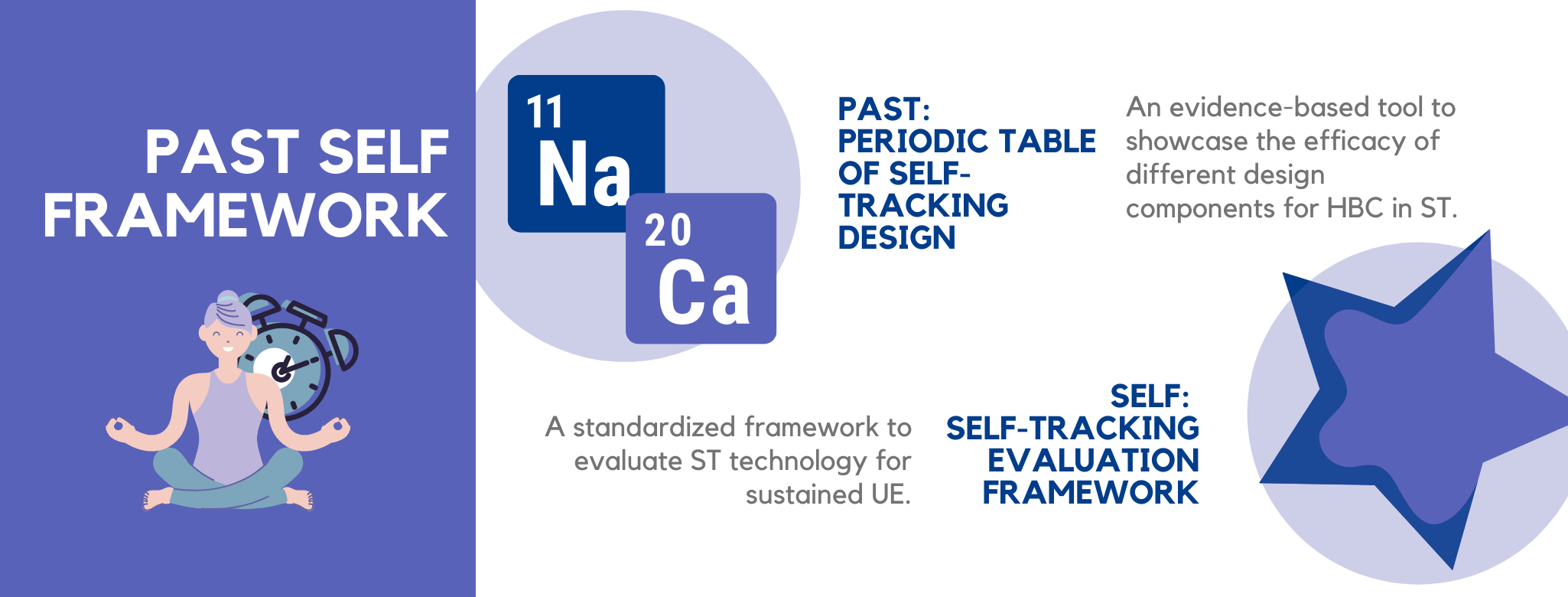}
  \caption{PAST SELF framework encapsulates the Design (PAST) and the Evaluation (SELF) Components.}
  \Description{PAST SELF framework encapsulates the Design (PAST) and the Evaluation (SELF) Components.}
  \label{fig:pastself}
\end{figure}

The main contribution of this paper is a large-scale systematic review of the literature related to ST technology. Through this review, we identify the lack of a common framework for designing and evaluating such technological interventions; hence, we propose PAST SELF, a novel, prescriptive framework for the design and evaluation of ST technologies' interfaces and system features (Figure~\ref{fig:pastself}). PAST SELF brings together principles of current practices and previous works in a formal and structured way. Thus, our contributions can be summarized as follows:
\begin{itemize}[leftmargin=*]
    \item \textbf{\textit{Systematic Review.}} 
    \begin{itemize}
        \item We explore the basic principles associated with ST technology for HBC (with a focus on PA) and UE and identify key challenges, limitations, and open questions that motivate our work (Section~\ref{background})
        \item We present our methodology for conducting a systematic review of the related literature, based on which we pose this article's research questions (Section~\ref{methodology}).
        \item We analyze included works and present a synthesis of the results related to interventions, experimental setups, theoretical frameworks, and publication details (Section~\ref{results}).
        \item We proceed to an in-depth study of the system design elements, namely interface components and system features, and how they affect the effectiveness of ST interventions, as well as the PA- and UE-related evaluation metrics most commonly used in ST research (Section~\ref{discussion}).
    \end{itemize}
    \item \textbf{\textit{PAST SELF.}} 
    \begin{itemize}
        \item We propose a framework to systematically classify and evaluate methods and experimental results in ST technology.
        \rev{Specifically, PAST SELF is a conceptual, prescriptive framework, in the sense that it guides system design and evaluation by transforming raw data into a set of abstractions and guidelines.} The PAST SELF framework consists of two components: The \textit{Design Component} (Section~\ref{periodicTable}) introduces the \textbf{P}eriodic T\textbf{a}ble of \textbf{S}elf-\textbf{T}racking Design (PAST), and enables ST developers and researchers to identify the most effective software and interface design elements for ST technology based on previous interventions. The \textit{Evaluation Component} (Section~\ref{self}) provides practitioners with a standardized way of measuring PA and UE to evaluate the effectiveness of their system over time, by introducing the \textbf{S}elf-Tracking \textbf{E}va\textbf{l}uation \textbf{F}ramework (SELF).
        \item We make our corpus of primary works that aggregates detailed information about the methods and metrics used in each of them, as well as numerical results of their experimental findings, publicly available as an open dataset \cite{sofia_yfantidou_2020_4063377}. Moreover, we provide an online, interactive tool\footnote{"PAST SELF Framework Tool." \url{https://syfantid.github.io/past-framework-visualization/}. Accessed 4 Jan. 2021.} for the visualization of the PAST component of our framework, \rev{as well as its source code for reusability purposes \cite{tom_valk_2021_4432162}}. 
        We deem that the corpus dataset (which is open to contributions) and the visualization tool can further facilitate ST researchers and practitioners beyond this review's scope.
    \end{itemize}
\end{itemize}

\section{Background}\label{background}
In this section, we initially provide an overview of the HBC and UE technology aspects, which are identified as crucial aspects for the ST technology's success, as seen in Section~\ref{introduction}. We provide a formal background on these aspects valuable for ST practitioners and researchers. Moreover, we discuss relevant past reviews and meta-analyses to reveal critical challenges and limitations, as well as the novelty of the present work.

\subsection{Health Behavior Change (HBC)} \label{hbc}
In the context of public health, HBC refers to ``efforts put in place to change people's habits and attitudes, to prevent disease'' \cite{world2010global}. HBC efforts can be aimed at different levels, including individual, organizational, community, and population levels, and there exists an interaction between them.
HBC programs usually utilize behavioral change theories at different levels. Individual and interpersonal theories are frequently encountered in the field of ST technology, including, but not limited to, the Transtheoretical Model of Health Behavior Change \cite{prochaska1997transtheoretical}, and the Self-Determination Theory \cite{deci2008self}. To better understand the theories implemented in ST technology, we provide a qualitative analysis of theoretical frameworks in ST in Section~\ref{results}. 
\begin{table}[htb!]
\resizebox{\linewidth}{!}{%
\begin{tabular}{ll}
\hline
\multicolumn{2}{c}{\textbf{Descriptions of PSD model strategies}}                                                                                                                                      \\ \hline
\multicolumn{2}{l}{\textbf{Primary task support}}                                                                                                                                                      \\ \hline
Reduction                & The system has to decrease the effort and strain users consume when doing their target behavior by reducing complex behaviors into simple and easy tasks for users.       \\
Tunneling                & The system has to guide users in the attitude change process or experience by providing opportunities for action performance that makes the user nearer to the target behavior. \\
Tailoring                & The system offers tailored information for its user group according to their interests, needs, personality, or other factors related to the user group.               \\
Personalization          & The system has to provide personalized content and customized services for users.                                                                                           \\
Self-monitoring          & The system has to give means for users to track and monitor their performance, progress, or status in accomplishing their goals.                                            \\
Simulation               & The system needs to give means for observing and noticing the connection between the cause and effect of users' behavior.                                                   \\
Rehearsal                & The system must deliver means for rehearsing a target behavior.                                                                                                             \\ \hline
\multicolumn{2}{l}{\textbf{Dialogue support}}                                                                                                                                                          \\ \hline
Praise                   & The system has to deliver praise through images, symbols, words, videos, or sounds as an approach to give user feedback information regarding his/her behavior.             \\
Rewards                  & The system should offer virtual rewards for users to provide credit for doing the target behavior.                                                                          \\
Reminders                & The system has to remind users to perform their target behavior while using the system.                                                                                     \\
Suggestion               & The system has to suggest ways that users can achieve the target behavior and maintain performing behavior during the use of the system.                                    \\
Similarity               & The system must imitate its users in some particular manner, so the system should remind the users of themselves in a meaningful way.                                       \\
Liking                   & The system should be visually attractive and contain a look and feel that meets its users' desires and appeal.                                                           \\
Social role              & The system has to adopt a social role by supporting the communication between users and the system's specialists.                                                           \\ \hline
\multicolumn{2}{l}{\textbf{System credibility support}}                                                                                                                                                \\ \hline
Trustworthiness          & The system has to give truthful, fair, reasonable, and unbiased information.                                                                                                \\
Expertise                & The system has to offer information displaying experience, knowledge, and competence.                                                                                       \\
Surface credibility      & The system must have a competent look and feel that portrays system credibility based on an initial assessment.                                                             \\
Real-world feel          & The system must give information about the organization or the real individuals behind its content and services.                                                           \\
Authority                & The system should refer to people in the role of authority.                                                                                                                 \\
Third-party endorsements & The system should deliver endorsements from well-known and respected sources.                                                                                               \\
Verifiability            & The system has to give means to investigate the accuracy of the system content through external sources.                                                                    \\ \hline
\multicolumn{2}{l}{\textbf{Social support}}                                                                                                                                                    \\ \hline
Social learning          & The system has to give a user the ability to observe other users and their performance outcomes while they are doing their target behavior.                                 \\
Social comparison        & The system should enable users to compare their performance with other users' performance.                                                                                  \\
Normative influence      & The system has to have a feature for gathering together individuals that have identical objectives and let them feel norms.                                                 \\
Social facilitation      & The system should enable a user to discern other users who are performing the target behavior along with him/her.                                                           \\
Cooperation              & The system should offer the opportunity for a user to cooperate with other users to achieve the target behavior goal.                                                       \\
Competition              & The system should allow a user to compete with other users. In the competition principle, there is a chance for winning or losing a race.                                   \\
Recognition              & The system has to offer public recognition (e.g., ranking) for users who do their target behavior.               \end{tabular}%
}
\caption{The strategies of the Persuasive System Design (PSD) Model \cite{oinas2008systematic}.\label{tab:psd}}
\vspace{-6mm}
\end{table}

However, behavioral change theories do not provide specific details on how their theoretical components could be translated into a real-world HBC system, such as ST technology. Thus, the interpretation is up to the ST practitioner. To bridge this gap, \citet{oinas2008systematic} have proposed a computer-science-based framework for Persuasive Systems Design (PSD), which is theory-creating by its nature. PSD is widely adopted and appreciated as a \rev{model} that describes the content and software functionality required in a Behavior Change (BC) product or service. For these reasons, we organize and present the persuasive strategies utilized by the included studies (Sections \ref{pts} to \ref{o}), and we build the PAST component of our framework (Section~\ref{periodicTable}) based on the generic PSD principles. In particular, the PSD framework defines four categories of strategies, namely, \textit{primary task support}, \textit{dialogue support}, \textit{credibility}, and \textit{social support}; each category has seven sub-groups within. Table~\ref{tab:psd} summarizes the PSD model strategies for BC. As a final note, we would like to mention that the PSD framework per se provides no information about each of its elements' effectiveness in achieving HBC, which we discuss below and address in our framework. Specifically, in the following sections, we present and discuss results of previous ST studies on the effectiveness of different system interaction components (Sections \ref{pts} to \ref{o}) and propose a way to formalize the quantification of their efficiency in our PAST SELF Framework (Section~\ref{periodicTable}).

\subsection{User Engagement (UE)} \label{ue}
For HBC to translate into population health, it must be maintained over the long run \rev{regardless of the use of technological interventions. It is important to note that attrition of a single ST technology does not equal abandonment of the desired behavior. There are two categories of lapses when it comes to ST technology: short-term and long-term lapses, and the latter may lead to attrition \cite{Epstein2015b}. However, the causes behind such lapses are varied, ranging from the high cost of collecting and integrating, or having and sharing data, to changed life circumstances, and accomplished or alternative goals and contexts \cite{Epstein2016c}. Additionally, users might choose to switch between ST systems to match their dynamic needs and health goals; hence abandoning one system does not necessarily mean abandoning ST or the effort for HBC \cite{rooksby2014personal}. Nevertheless, in the context of ST technology, HBC systems need to measure, increase, or sustain UE to drive users towards HBC. If abandoned early, ST technology can evidently not accompany the users in their HBC journey, which may or may not continue independently. A clearer definition of UE will enable us to set the foundations for its quantification.} 

\citet{attfield2011towards} define UE as "the quality of the user experience that emphasizes the positive aspects of the interaction, particularly the phenomena associated with wanting to use a technological resource longer and frequently". UE is vital to measure because it can quantify if the user interacts with the system successfully or not, avoiding higher attrition rates, a common problem of ST technology, as seen in Section~\ref{introduction}.
We adopt a three-faceted view of UE: the emotional aspect, the cognitive aspect, and the behavioral aspect of UE \cite{lalmas2014measuring,pilotti2017factors}. The {\it emotional aspect} refers to the user's feelings and state of mind regarding the system and is usually measured through self-reports. The {\it cognitive aspect} refers to the user's physical reaction to the system (e.g., eye gaze, bodily response). It is usually measured through physiological measurements, such as body temperature and heart rate measurements. Finally, the {\it behavioral aspect} refers to the user's behavioral response to the system (e.g., frequency of visits, duration) and is usually measured by analytics, such as analytics on usage logs. Each aspect captures UE's different characteristics, and a combination of all aspects offers a holistic view. Several UE metrics that fall under these aspects refer mostly to desktop computing or, in some cases, mobile computing, where measurements, such as eye-tracking or mouse-tracking, are still possible. Hence, in our work, we screen the included studies for elements related to the emotional, cognitive and behavioral aspects of UE by adapting and refining the generic methodology of \citet{lalmas2014measuring} to the ubiquitous ST technology domain. This refinement leads to the creation of the SELF component of our framework, as discussed in Section~\ref{self}. 

\subsection{\rev{Related Work in ST Technology Research for HBC and UE: Limitations \& Open Questions}} \label{lr}
Various studies have examined and evaluated the effectiveness of ubiquitous interventions for HBC with a focus on reducing sedentary behavior or increasing PA for individuals. Some researchers have conducted reviews that include interventions targeted to samples with specific characteristics, such as \rev{age group \cite{vargemidis2020wearable,Gerling2020}}, race \cite{muller2016effectiveness}, mental health issues \cite{Murnane2018}, or prior experience with ST \cite{Rapp2016}. Others focus on examining the effects of specific behavior change techniques, such as incentives \cite{strohacker2014impact}, personalization \cite{monteiro2019personalization}, \rev{social sharing \cite{Epstein2015c}, data summarization \cite{harris2021framework}, or technological advancements \cite{thieme2020machine}} on the activity levels of individuals. While these studies provide valuable knowledge for future research in tailored interventions, it is evident that due to the strict inclusion criteria, these reviews suffer from a limited number of primary studies, and their results might not be generalizable to the whole population. At the same time, some reviews do not focus on ubiquitous technology solutions, failing to capture the requirements of designing and evaluating modern ST technology. Also, due to the subject's multidisciplinarity, multiple reviews from different domains (e.g., medicine, psychology) seem to neglect recommendations for designing and evaluating ST technology, thus failing to bridge the gap between theoretical foundations and practice.

\rev{On the other hand, reviews that explore the design space for HBC offer insights for extended explorations targeted by our work. For instance, researchers in persuasive computing focus on using ST technology to persuade people to change their health behaviors. \citet{matthews2016persuasive} and \citet{10.3389/fcomp.2020.00019} have conducted systematic reviews of 80 and 20 papers respectively to assess the effectiveness of mobile phone-based interventions in encouraging PA and identify research trends in the area. However, they neither assess the effectiveness of individual intervention components nor propose a comprehensive methodology for the overall evaluation of such interventions. To address the first limitation, \citet{aldenaini2020trends} have expanded their initial work by publishing a second systematic review of 170 papers, where they have evaluated the effectiveness of individual intervention components in promoting PA. Similar to our work, they categorize intervention components under the PSD framework and report the success rates per technique. However, due to the static nature of their report, it is cumbersome for the reader to assess the effectiveness of such components for a distinct sample population, intervention duration, or sample size. Additionally, none of the aforementioned works culminates their review into a prescriptive, end-to-end framework, such as ``PAST SELF'' for designing and evaluating ST technology.}

\rev{To encounter such frameworks, we need to explore HCI research in the field of ST technologies for HBC. Several HCI user studies have culminated in frameworks, as well as models and guidelines, for designing successful HBC technologies. Specifically, \citet{Li2010} have proposed the widely used stage-based model of personal informatics systems composed of five stages (preparation, collection, integration, reflection, and action) that describe the user's experience with ST technology through time. Similarly, \citet{Epstein2015b} have introduced the lived informatics model of personal informatics by surveying and interviewing past and present trackers of PA, finances, and location regarding their experiences. Both models are fundamental in HCI literature for HBC and provide valuable insights for ST technology design. However, their focus is inherently different from our framework's, which aims to provide researchers and HCI designers with actionable insights on translating theory into practice (prescriptive framework) rather than high-level recommendations (explanatory framework). Also, \citet{Elsden2016} have proposed ``Quantified Past'', a model that seeks to explore how to design for long-term use, while \citet{kumar2020mobile} conducted a literature review on mobile and wearable sensing frameworks for mHealth.
Nevertheless, all of the frameworks and models mentioned above are conceptually different, but of no less importance than the one proposed in this paper, in a sense that they are a result of empirical HCI research and not based on the systematic review and synthesis of years of literature in the field.}

\rev{Evidently, though, HCI researchers have also conducted several reviews in the field of ST technology for HBC. \citet{Ayobi2016} have published a review of 20 papers, identifying and characterizing three streams of research in personal informatics– psychological, phenomenological, and humanistic. 
Similarly, in their work, \citet{kersten2017personal} have reported promising insights and methodological pitfalls drawn from 24 empirical studies utilizing ST technology. While these reviews introduce distinct research directions, they do not provide any actionable information regarding designing and evaluating ST technology. Additionally, they are limited in scope, since they draw their primary studies solely from a computer science digital library (despite the interdisciplinarity of the domain).  More recently, \citet{Epstein2020} have published an exhaustive mapping review of the personal informatics literature, where they seek to answer questions related, but not limited, to identifying personal informatics sub-domains, tracking motivations, challenges and ethical concerns, and types of research contributions. While this mapping review is a significant contribution to the ST literature, its scope is orthogonal to this systematic review. Mapping reviews aim to summarize the range of findings on a research topic at a high-level. By comparison, systematic reviews synthesize and summarize those findings, for example, into comprehensive guidelines or prescriptive frameworks, such as PAST SELF.}

Summing up, to the best of our knowledge, there are no systematic reviews that provide an evidence-based, end-to-end solution for designing and evaluating ST technology based on synthesizing, correlating, and adapting past studies' results. Moreover, no review has bridged the gap between HBC and UE quantification in the ST domain as far as we are aware. The limited reviews that refer to UE focus primarily on UE on websites or generic software \cite{sigerson2018scales}, rather than on HBC technology.
Hence, the current work aims to bridge these gaps by following a formal research methodology, which we present in the following section.

\section{Research Methodology}\label{methodology}
The current study follows an established methodology to determine how researchers have approached the design and evaluation of ubiquitous ST technology solutions for HBC and UE. Precisely, we follow a systematic methodology to ensure the quality of included studies and limit the initial number of articles, based on the guidelines introduced by Kitchenham's \cite{kitchenham2007guidelines} widely recognized protocol for conducting a systematic review.

Based on this protocol, we initially identified the need for a systematic review. In Section~\ref{lr}, we gave an overview of the literature reviews and frameworks surrounding the use of ST technology in HBC interventions. However, none of them focused on this article's objective, namely identifying, categorizing, and presenting best practices for designing and evaluating ST technology for HBC and UE \rev{with a focus on PA}. At the same time, we could not locate any similar work. Based on the Kitchenham methodology criteria and the research gaps discussed in Sections~\ref{introduction} and \ref{lr}, we specify the following five research questions that drive our review. We believe that these questions provide researchers in the domain or general stakeholders with a comprehensive view of UE and HBC strategies in ubiquitous ST technology, with the ultimate purpose of increasing PA levels.
\begin{description}
    \item[RQ1] How many research studies exist that address issues related to sustained UE and HBC in ST technology? 
    \item[RQ2] Which set-ups have been used (e.g., sample size, intervention duration) for HBC and UE experiments?  
    \item[RQ3] Which are the theoretical frameworks that have been used to increase PA and sustain UE? 
    \item[RQ4] Which are the most effective HCI design strategies for ST technology (interface and system components and functionalities) that have been used to achieve HBC and sustain UE? 
    \item[RQ5] How can we measure PA and evaluate UE in ST technology? 
\end{description}

To locate the papers that would help us answer the questions above, we chose to perform a broad, automated search in digital libraries, focusing on articles that have been published in journals and conferences for the last 18 years (between 2004 and 2022). \rev{Early research works in the field of ubiquitous technology for HBC with a focus on PA (see related reviews \cite{Epstein2020,consolvo2009goal}), include Lin et al.'s Fish'n'Steps (2006) \cite{Lin2006}, and Consolvo et al.'s Houston (2006) \cite{consolvo2006design} and UbiFit Garden (2008) \cite{Consolvo2008}.} Additionally, commercial wearable technology for ST (e.g., Nike + iPod Sport Kit, Fitbit) emerged in the late 2000s. \rev{Hence, our research time range is chosen based on the aforementioned scientific and industrial advancements.} As our sources, we utilized Google Scholar, Scopus, IEEE Xplore, and Web of Science digital libraries for their coverage and accessibility or the quality of their results. To identify appropriate search terms combined with Boolean operators, we followed the guidelines of \citet{spanos2016impact}. According to the authors, "the determination of search terms is an iterative procedure starting with trial searches using different search terms, considering an initial set of articles that is already known to belong to the research field of the systematic review". The procedure of determining search terms ends "when the initial set of already known articles is found by the search". Following this procedure led us to the search query below:
\begin{tcolorbox}
("mobile application" OR "mobile phone" OR smartphone OR "digital coach" OR "digital trainer" OR wearable OR "activity tracker" OR "self-tracking devices" OR smartwatch)\\AND\\("user study" OR "persuasive technology" OR "user engagement" OR "user motivation")\\AND\\(fitness OR "physical activity" OR exercise)
\end{tcolorbox}
To ensure the high quality and relevance of the included papers, we define appropriate inclusion and exclusion criteria which help us determine the final sample of articles:
\begin{description}
    \item[Inclusion Criteria]\hfill
    \begin{enumerate}
        \item Articles published in a peer-reviewed academic journal or proceedings from an international scientific conference (NPR); 
        \item Articles published in English (NE);
        \item Articles that include at least one user experiment, such as an intervention, a pilot study, or a longitudinal study (NI);
	\item Articles that utilize at least one ubiquitous device, such as a mobile or a wearable device (NUI);
        \item Articles that include a quantitative assessment of the intervention's effect, either in terms of altered PA levels or user engagement (NM).
    \end{enumerate} 
    \item[Exclusion Criteria]\hfill
    \begin{enumerate}
        \item Articles that discuss HBC and UE in different domains, such as Marketing or Behavioral Economics (DD);
        \item Articles that discuss other forms of HBC, such as nutrition monitoring, stress monitoring, or smoking cessation, rather than physical activity (NPAO);
        \item Articles that refer to interventions that utilize a subjective assessment of activity levels, e.g., utilizing surveys measuring the perceived level of activity or manual data entry (NS);
        \item Articles that use out-of-the-box products, \rev{without any additional intervention components}, solely for monitoring performance (OP);
        \item Articles that do not describe the ubiquitous ST technology's design features in the description of the user experiment (NDF);
	\item Articles published in different outlets but referring to the same user experiment (D).
    \end{enumerate}
\end{description}
\begin{figure}[htb!]
  \centering
  \includegraphics[width=.95\linewidth]{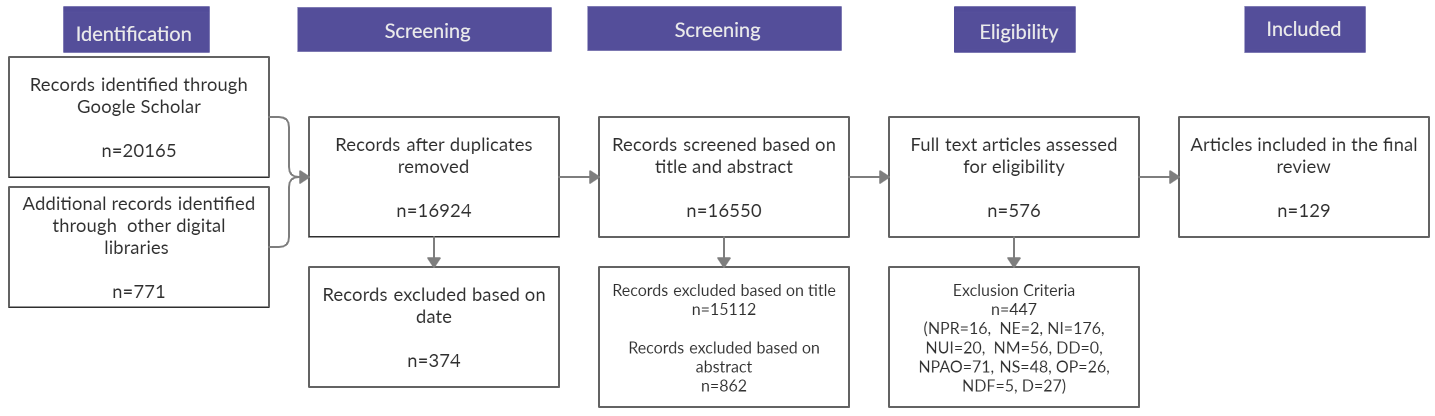}
  \caption{PRISMA flow diagram of studies in the systematic review.}
  \Description{A PRISMA flow diagram of studies in the systematic review. Out of 20165 returned records, 576 were admitted for full-text assessment, and 129 were included in the review.}
  \label{fig:PRISMA}
\end{figure}

Finally, the sequential execution of the steps above led to our review's final set of articles. Our search results from various digital libraries are depicted in Figure~\ref{fig:PRISMA}. Overall, we screened 16924 articles after duplicate elimination, of which we removed 374 based on date criteria (earlier than 2004), 15112 based on the title, and 862 based on the abstract. After carefully studying the remaining 576 articles, we excluded 447 articles based on our inclusion/exclusion criteria. Hence, 129 articles synthesized our final pool. We present the data features we extracted from the included articles and the results of their synthesis in the following sections.

\section{\rev{Quantitative Findings from the Systematic Review}}\label{results}
In this section, we present our quantitative findings based on the investigated articles. The section provides answers to our research questions RQ1, RQ2, and RQ3; namely, it presents the number and venues of published studies related to sustained UE and HBC in ST technology over the years, the experiment set-ups they have utilized, as well as the theoretical frameworks behind the intervention design. 

\begin{figure}[htb!]
  \centering
  \includegraphics[width=.75\linewidth]{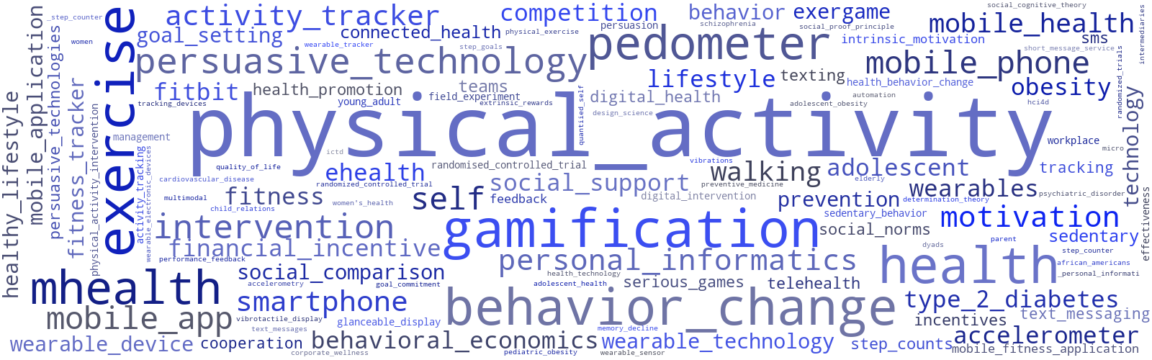}
  \caption{A word cloud of the keywords of our final article pool.\label{fig:wordcloud}}
  \Description{A word cloud of the keywords of our final article pool.}
\end{figure}
We first present a general overview of the scientific topics in the field of ST technology for HBC in Figure~\ref{fig:wordcloud}. The depicted word cloud is produced from the keywords in our article pool. It illustrates topics gathering scientific interest in the domain of ST technology for HBC. \rev{Not surprisingly}, "physical activity", "behavior change", and "mHealth", are some of the main topics of interest. It also highlights some of the design techniques utilized for HBC interventions (e.g., gamification, social support, competition) and the sampled populations' characteristics (e.g., adolescents, type-2 diabetes, obesity). Finally, it verifies the interdisciplinarity of the field, as discussed in previous sections, with terms emitting from Behavioral Economics, Medicine, Psychology, and Computer Science. 

\begin{figure}[htb!]
  \centering
  \includegraphics[width=.9\linewidth]{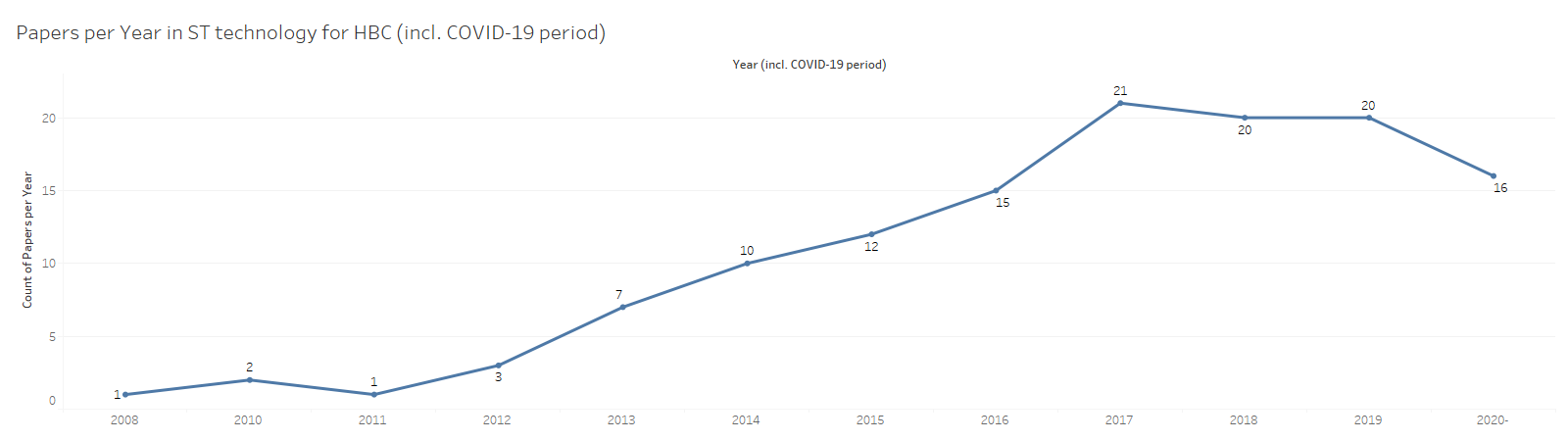}
  \caption{The trend of published articles per year in the domain of ST technology for HBC.\label{fig:years}}
  \Description{The trend of published articles per year in the domain of ST technology for HBC. We can see an increase in published articles over the years (with an exclusion of COVID-19 years).}
\end{figure}

\begin{figure}[htb!]
  \centering
  \includegraphics[width=\linewidth]{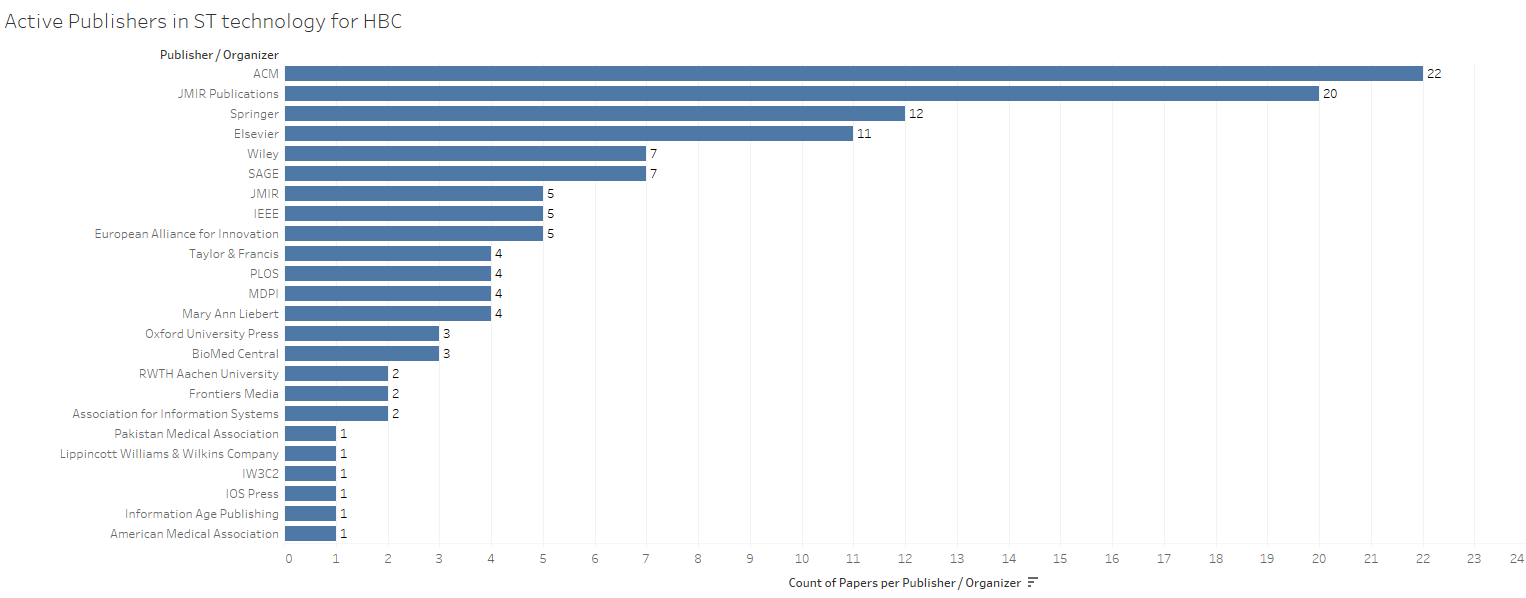}
  \caption{A bar plot of the most active publishers in the domain of ST technology for HBC.\label{fig:publishers}}
  \Description{A bar plot of the most active publishers and conference organizers in the domain of ST technology for HBC. ACM, JMIR Publications, and Springer hold the top-3 positions.}
\end{figure}

Regarding our research questions, we present the results in response to RQ1 in Figures~\ref{fig:years} and \ref{fig:publishers}, which summarize the number of publications per year related to ST and HBC technology and the number of publications per publisher or conference organizer, respectively. 

User interventions utilizing commercial or custom-made ST technology for HBC first appeared in 2008, which is reasonable since ST technology became commercially available in the late 2000s. Since then, the number of related publications has faced a steady increase, while it has more than doubled in three years (between 2014 and 2017). Note that Figure~\ref{fig:years} is only based upon our final article pool (129 articles), but our exploration of the non-eligible article pool (447 articles) follows a similar trend. \rev{It is important to note that there is a significant body of work in the field prior to 2008, which refers to artifact contributions or experimental technology that led to the development of advanced interventions later on. While these works are ground-breaking in the field, they are out of the scope of this literature review, which focuses on user interventions for HBC and UE, their components, and reported results.}
\begin{tcolorbox}
\textbf{Key Finding:} ST technology for HBC research is still in its infancy, but with growing interest over the years.
\end{tcolorbox}

ACM (22 articles), JMIR Publications (20 articles), and Springer (12 articles) hold the top-3 of publishers/conference organizers in the field (See Figure~\ref{fig:publishers}). The most popular journals for publishing articles regarding ST technology in HBC include JMIR mHealth and uHealth (15 articles), followed by PLoS ONE and the American Journal of Preventive Medicine (4 articles each). Similarly, the most popular conferences include the ACM CHI Conference on Human Factors in Computing Systems (6 articles), the ACM International Joint Conference on Pervasive and Ubiquitous Computing (UbiComp) (5 articles), and the EAI PervasiveHealth Conference (4 articles).
\begin{tcolorbox}
\textbf{Key Finding:} The interdisciplinarity of the domain is evident by its main publishing venues; medical and medical informatics journals hold the first spots for journals, while computer science conferences are the most popular.
\end{tcolorbox}

\begin{figure}[htb!]
  \centering
  \includegraphics[width=\linewidth]{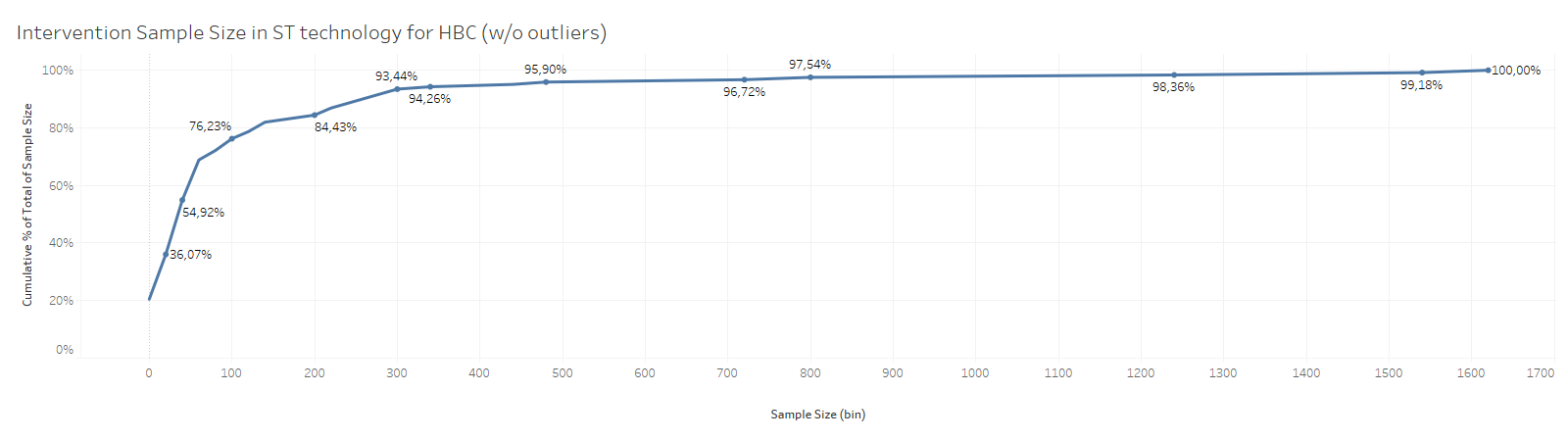}
  \caption{A cumulative percentage plot of intervention sample size. Three outlier sample sizes larger than 1700 subjects were excluded for presentation purposes. \label{fig:samplesize}}
  \Description{A cumulative percentage plot of intervention sample size. More than 75\% of the conducted experiments have a sample size of fewer than 100 users.}
\end{figure}
\begin{figure}[htb!]
  \centering
  \includegraphics[width=.9\linewidth]{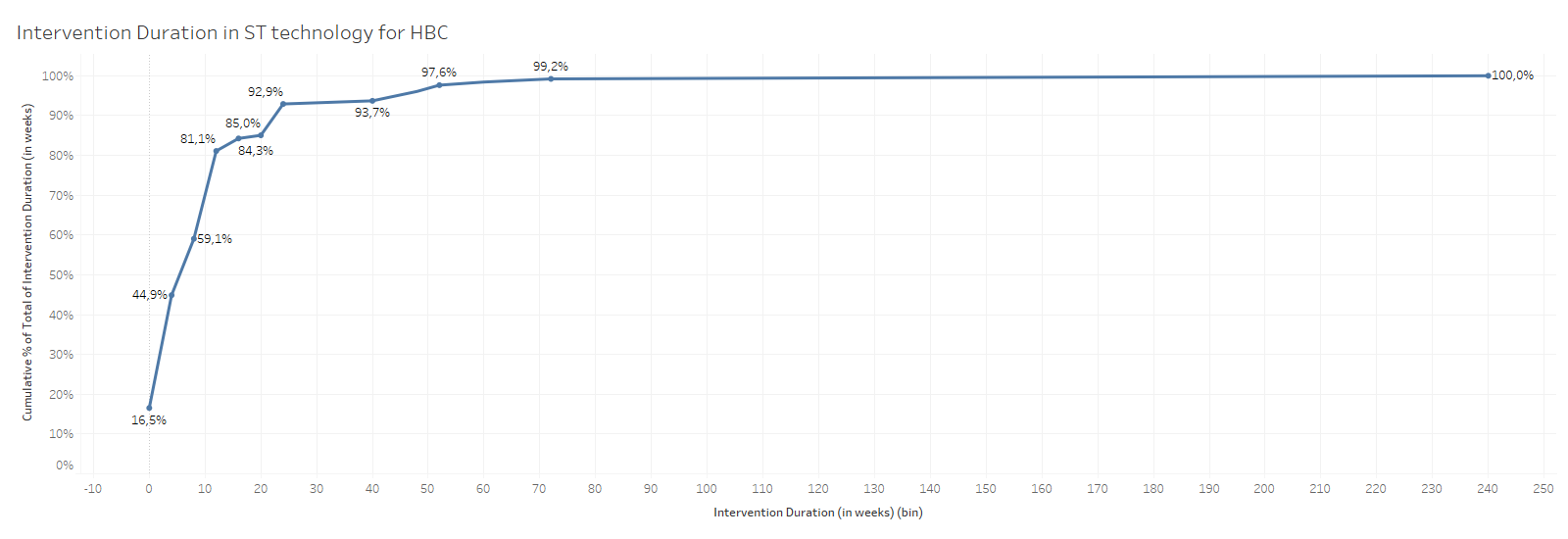}
  \caption{A cumulative percentage plot of intervention duration (in weeks).\label{fig:duration}}
  \Description{A cumulative percentage plot of intervention duration. Approximately half of the conducted experiments have a duration of less than a month.}
\end{figure}

Figures~\ref{fig:samplesize} and \ref{fig:duration} present the results in response to RQ2; namely, they outline the intervention set-ups for HBC utilizing ST technology. Figure~\ref{fig:samplesize} shows that more than three-quarters of all investigated studies have a sample size of fewer than 100 subjects, while more than a third have a sample size of less than 20. Similarly, in Figure~\ref{fig:duration}, we can see that 4 out of 5 interventions have a duration of fewer than three months, while almost half have a duration of less than a month. While such sample sizes and duration may be understandable in an experimental setting, small samples and short duration can negatively affect the generalizability of interventions' results \cite{faber2014sample}. Hence, these observations highlight the need for large-scale experiments and provide a fertile space for future experimentation with larger populations over extended periods.
\begin{tcolorbox}
\textbf{Key Finding:} The majority of interventions utilize a small sample size ($<100$ subjects) and short duration ($<1$ month), causing generalizability issues for the intervention outcomes and asking for large-scale interventions.
\end{tcolorbox}

The mean age of the sample population for all studies is 38 years old. The majority of the experiments refer to young adults (22\%), i.e., university students, or adults and middle-aged people (48\%), while fewer experiments focus on kids (5\%) and adolescents (13\%) or the elderly (12\%). However, the population worldwide is ageing\footnote{"Ageing | United Nations ." \url{https://www.un.org/en/sections/issues-depth/ageing/}. Accessed 15 Sep. 2020.}, meaning that we are experiencing a growth in the number and proportion of older people. Hence, ST technology should cater to the needs of this growing market segment, and future studies should focus on exploring the design space of ST devices for the elderly. \rev{At the same time, adolescents are active users of technology and also the most inactive population subgroup, with 3 in 4 adolescents (aged 11–17 years) not meeting the global recommendations for PA set by WHO \cite{world2019global}. Future research should be directed towards designing adolescent-oriented ST technology by prioritizing the enhancement of features that they can offer to this user subgroup. Researchers working with sensitive user groups, such as the elderly, kids, and adolescents, should always consider the ethical and societal implications behind technological intervention deployments with these cohorts, as discussed in Section~\ref{conclusions}.} 
\begin{tcolorbox}
\textbf{Key Finding:} While research currently focuses on the adult population, ST technology for the elderly or adolescents demonstrates a hidden potential for the future.
\end{tcolorbox}

\rev{Apart from the age criterion, a small percentage of studies have focused on populations with different characteristics regarding gender (<8\%), ethnicity (<3\%), physical health (<18\%), PA level (<8\%), employment status (<16\%), and mental health (<1\%)}. The most popular ST technologies utilized for the experiments involve Fitbit activity trackers (36 articles), ActiGraph accelerometers (12 articles), and Polar, Omron, and Jawbone activity trackers (4 articles each). Whenever an intervention focused on mobile ST technology, it usually utilized the mobile's built-in sensors and a custom-made application. Much fewer interventions incorporate powerful out-of-the-box APIs, such as the Google Fit SDK, or Apple's Core Motion API. \rev{We presume that the following two reasons have contributed to this situation. First, this could be because many of the researchers behind ST for HBC interventions do not have a computer science background and potentially lack the skill set to develop more technologically advanced interventions. More importantly, though, many of these technologies have only been made available recently and are still in development. For example, Apple's CoreMotion API was only released in 2013, meaning that researchers conducting interventions previously had to develop every component from scratch.} Nowadays, such APIs facilitate developers in creating ST technology interventions by giving them easy access to sensor data and users' real-time and historical PA data and integrating with popular ST devices.  
\begin{tcolorbox}
\textbf{Key Finding:} There is an abundance of existing, out-of-the-box APIs and ML libraries that can facilitate and enhance ST interventions development. However, up to now, studies have been limited to exploiting only a fraction of these capabilities. 
\end{tcolorbox}

\begin{figure}[htb!]
  \centering
  \includegraphics[width=\linewidth]{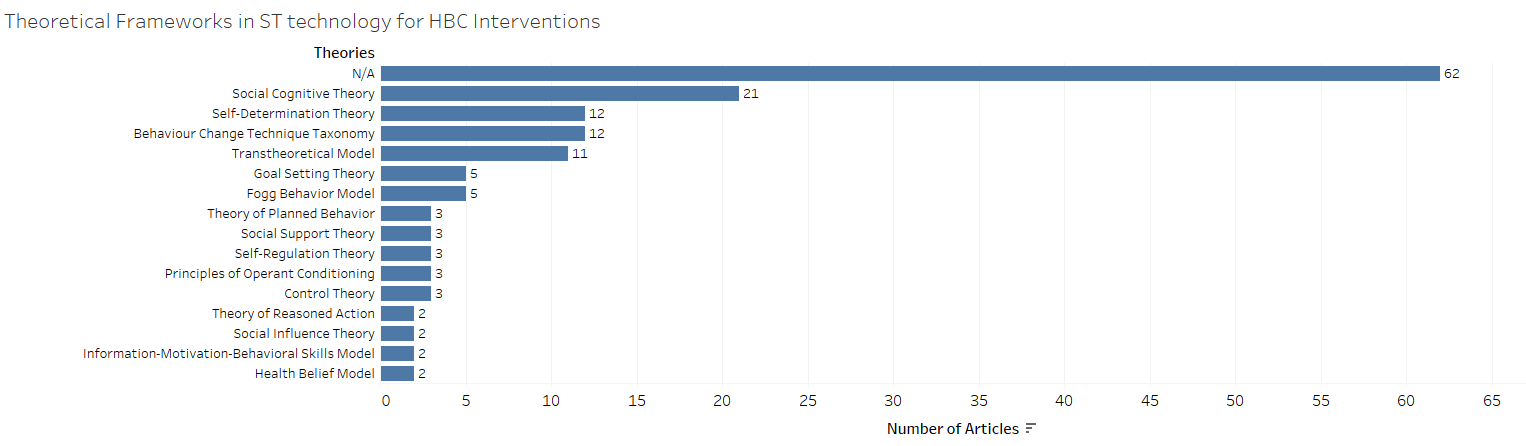}
  \caption{The theoretical frameworks utilized for intervention design with more than one occurrences in the investigated articles.\label{fig:theories}}
  \Description{The theoretical frameworks utilized for intervention design with more than one occurrences in the investigated articles. Social Cognitive Theory, Behaviour Change Technique Taxonomy, and the Behaviour Change Technique Taxonomy are the top-3 theories used.}
\end{figure}
Figure~\ref{fig:theories} gives us an answer to RQ3; namely, it outlines the theoretical frameworks utilized to support HBC interventions. Notice that almost half of the investigated articles (specifically 48\%) either did not utilize any theoretical framework or did not clearly identify one in their full text, limiting their potential impact. This absence is partly driven by the plethora of overlapping behavior change theories and related strategies and the lack of domain expertise and interdisciplinary research \cite{pinder2018digital}. \rev{However, BC theories can positively affect ST technology by informing design, guiding evaluation, and inspiring alternative experimental designs \cite{Hekler2013}}. From those articles that did mention a psychological theory, we can see that the four most popular theories include the Social Cognitive Theory (21 articles) \cite{ratten2007social}, the Behaviour Change Technique Taxonomy \cite{michie2013behavior} and the Self-Determination Theory \cite{deci2008self} (12 articles each), and the Transtheoretical Model (11 articles) \cite{prochaska1997transtheoretical}. While these theories mostly explain human behavior rather than provide methods for designing and evaluating ST technology, they work as foundations behind ST technology's proposed interventions. The PAST SELF framework considers several interventions, backed by different theories, bringing their insights together to provide a common framework for designing and evaluating ST technology for HBC and UE.
\begin{tcolorbox}
\textbf{Key Finding:} There is no common framework for ST interventions. Also, almost half of the investigated articles were not based on any theoretical framework or did not clearly identify one in their full text.
\end{tcolorbox}

To sum up, our key findings showcase that ST technology research for HBC is still in its infancy with growing interest from various disciplines (RQ1). It is a field with several unexplored facets with great potential for research, e.g., conducting experiments with larger sample sizes, longer duration, and varied population characteristics (RQ2). Finally, the lack of utilization of development tools and theoretical frameworks (RQ3) highlights the need for a standardized, robust, and extensible framework for the design and evaluation of ST systems for HBC and UE; this is the gap that the PAST SELF framework we propose (Section~\ref{discussion}) aims to fill.

\section{From a Systematic Review to the PAST SELF Framework}\label{discussion}
In Section~\ref{results}, we offer an evidence-based response to three of our research questions, RQ1, RQ2, and RQ3. This section responds to RQ4 and RQ5 through the thorough study of the 129 investigated articles and the synthesis of the outcomes towards establishing a systematic framework. In the following subsections, we present an analysis of the most common HCI design (Section~\ref{design}) and evaluation (Section~\ref{self}) techniques encountered in the included articles, providing answers to RQ4 and RQ5, respectively. 

Additionally, we bring together past results and current practices in the form of the PAST SELF Framework. PAST SELF is an evidence-based framework for identifying, categorizing, and presenting ST technology's most common design and evaluation elements for HBC and UE. Specifically, in Section~\ref{periodicTable} we propose the design component of the framework, the \textbf{P}eriodic T\textbf{a}ble of \textbf{S}elf-\textbf{T}racking Design (PAST), and in Section~\ref{self} we propose the evaluation component, the \textbf{S}elf-Tracking \textbf{E}va\textbf{l}uation \textbf{F}ramework (SELF). Finally, in Section~\ref{usecases}, we demonstrate how researchers and practitioners in the field can apply and benefit from the PAST SELF framework through a use case scenario. PAST SELF aims to help researchers learn from previous experiences, identify best practices, facilitate and accelerate their current research, and avoid common pitfalls.  

\subsection{Designing for Health Behavior Change \& the PAST Component}\label{design}
This section discusses how the general-purpose PSD persuasive techniques have been used in practice in prior research to design ST interfaces and system features. To achieve this and provide an answer to RQ4, we manually extract PSD elements from the investigated papers and discuss them in Sections~\ref{pts} to \ref{o}. 
Finally, in Section~\ref{periodicTable}, we bring together the knowledge extracted from the investigated literature, presenting the PAST Component of our PAST SELF Framework, which gives us an evidence-based insight into the effectiveness of different persuasive techniques for designing successful ST technology.

\subsubsection{Primary Task Support}\label{pts}\hfill\\
In the PSD Framework, Primary Task Support includes seven persuasive techniques: Self-monitoring, Personalization, Reduction, Tailoring, Tunneling, Simulation, and Rehearsal (See Table~\ref{tab:psd}).

In the investigated papers, self-monitoring usually takes the form of online visual, textual, audio or haptic feedback on a mobile application dashboard, a watch face or application, a phone background, a push notification, a web application dashboard, a public display, or another wearable device interface. It can also take the form of offline feedback, such as e-mail or SMS communication. The feedback can be both real-time and analytical, even though the latter is less common in the investigated literature. Real-time, textual performance feedback, such as step count, is the most common information provided to the users as an indication of their daily PA.  It comes as no surprise that self-monitoring is the most common PSD technique encountered in the literature, since recording and visualizing the users' data is by definition the purpose of ST technology. 

\textbf{Personalization} is implemented as personalized goals based on past PA \cite{Mutsuddi2012c,Bickmore2013,King2013,Washington2014a,AlAyubi2014,Zuckerman2014a,Arigo2015,Poirier2016a,Gouveia2016b,Mendoza2017,Chen2017,Harkins2017,Chung2017a,Korinek2018,Bianchi-Hayes2018b,Simons2018d,Zhou2018c,Gremaud2018b,Kooiman2018b,Choi2019b,Hochsmann2019e,Middelweerd2020b,ek2020effectiveness,larsen2020feasibility,damschroder2020effect,koontz2021increasing}, automatic identification of psychological state \cite{Leinonen2017b}, personalized exercise plans based on user goals, preferences or past behaviors \cite{Pilloni2013,Rabbi2015e,Lee2015b,Rabbi2015d,Boratto2017,Cadmus-Bertram2019b,Hochsmann2019e,stephens2022feasibility}, or generally personalized content based on user's habits, time of day, location, weather, or preferences \cite{Gouveia2014,Rabbi2015e,Agboola2016c,Cambo2017,VanDantzig2018c,Yoon2018b,Klasnja2019,Gell2020a,Robinson2019, Michael2020,Ding2016,wang2021smartphone}. Specifically, to set new, personalized goals, the system learns about the user's activity levels based on their past PA (usually daily step counts) and then recommends a more challenging daily step goal. The goal has to be challenging enough to motivate the user to increase their PA but not unrealistically demanding to avoid demotivating the user. Also, it has to be in line with the recommended guidelines. Similarly, personalized content, such as exercise plans, aims to contextualize PA in the user's lifestyle to make it easier to perform the target behavior, namely sufficient PA. Promoting active transportation, e.g., walking to work over driving, is an example of a personalized, contextualized suggestion. Note that, according to previous research \cite{li2012using}, associations between physical activity and contextual information help users become more aware of the factors that affect their PA levels. Finally, the application of Machine Learning (ML) for ST personalization is very promising and still at its early stages. Different users have different patterns of usage \cite{Rapp2018a} and future work should focus more on learning the users' needs, wants, pain points, and habits based on their past data to automatically adjust the systems' features to the individual user and maximize its results.

\textbf{Reduction} is achieved through multiple simultaneous goals \cite{Consolvo2008d,Foster2010a,Gouveia2014,Kramer2019,Chen2014b,Chen2016,Gouveia2015a,Verbeek2012a}, graded goals or difficulty levels \cite{King2013,Chen2016,Garde2015,Lee2015b,Rabbi2015d,Blackman2015b,Verbeek2012a,Patel2017a,Zhao2017b,Morrison2017,Tong2017b,Tu2019b,fuemmeler2020mila,nicholas2021development,koontz2021increasing}, contextualization of PA into everyday life \cite{Mutsuddi2012c,Finkelstein2015d,Martin2015b,Gouveia2015a,Wernbacher2020,wang2021smartphone}, exercise planning functionality or guidance \cite{Pilloni2013,Arigo2015,Blackman2015b,Lee2016,Verbeek2012a,Boratto2017,Cambo2017,Alsaqer2017,Chung2017a,Pope2018b,Hochsmann2019e,Klasnja2019,Galy2019b,Robinson2019,Michael2020,nicholas2021development,stephens2022feasibility}, social network integration \cite{AlAyubi2014,Chung2017a}, or feedback on progress towards goal accomplishment \cite{Verbeek2012a,Lyons2017b,Leinonen2017b,Morrison2017,VanDantzig2018c,Altmeyer2018b,Esakia2018c,Elliott2019b,Kim2018c,Zhang2019a}. The idea behind reduction is that the complexity of performing PA (time, physical and mental effort, and persistence required) can be decreased by modularizing PA and providing users with step-by-step guidance along the way. For instance, by providing graded, simultaneous step goals, the system gives even less active users (e.g., 6000 daily steps) the satisfaction of goal accomplishment to motivate them to perform better in the future (e.g., 8000 daily steps or even the recommended 10000 steps). On a different note, systems that offer on demand exercise plans of different duration, PA type and PA level, facilitate even the most novice users in the process of performing PA.

\textbf{Tailoring} takes the form of personalized information based on psychological profiles \cite{Mutsuddi2012c,Hebden2014a,Agboola2016c,Chen2017,Leinonen2017b,Middelweerd2020b,yuasa2022estimation}, gender \cite{Hebden2014a,Simons2018d,Schafer2018}, age \cite{Alsaqer2017,Bianchi-Hayes2018b,Galy2019b}, professional occupation \cite{Simons2018d,VanDantzig2018c,Middelweerd2020b}, interests \cite{Hochsmann2019e}, health status \cite{Pope2018b,Gell2020a,fuemmeler2020mila}, or season \cite{Kramer2017b}, as well as tailored motivational messages \cite{park2021mobile,koontz2021increasing}, and exercise plans based on user experience \cite{Pilloni2013,Gremaud2018b,nicholas2021development} or established guidelines \cite{Vathsangam2014}. For example, cancer survivors require different PA content than healthy adults, elderly with limited tech skills require different interfaces than the tech-savvy kids of today, exercise buffs need more demanding exercise plans than novices, and people of different ethnicities might have different habits in terms of PA that the system should take into consideration. \rev{Note that Tailoring differs from Personalization in that it caters for and adapts to the preferences of a user subgroup rather than an individual user.} Similar to Personalization, Tailoring ST technology has great potential in the age of ML. Future researchers can train inclusive models for sample populations with different characteristics to build more inclusive ST technology that caters to different user groups' needs.

\textbf{Tunneling} is realized through information provision and PA recommendation pairing \cite{FesslerMichaelB.;RudelLawrenceL.;Brown2008a,McNab2012b,larsen2020feasibility}, reminders and goal-setting pairing \cite{AlAyubi2014,Liu2016,Polgreen2018b,Robinson2019,nicholas2021development}, reminders and PA recommendation pairing \cite{Wernbacher2020}, goal-setting and feedback loop pairing \cite{nicholas2021development}, context or goal identification and PA recommendation pairing \cite{Rabbi2015e,Lee2015b}, gradual goal adjustments \cite{FesslerMichaelB.;RudelLawrenceL.;Brown2008a,Bickmore2013,ek2020effectiveness,stephens2022feasibility}, graded rewards availability \cite{Chen2014b,Zhao2017b}, or step-by-step guided PA routines \cite{Boratto2017,Cambo2017,Paul2017}. The concept behind tunneling is transforming PA, or any related target behavior, into a step-by-step process that the user can follow. For example, based on the investigated literature, users are less likely to set their own goals autonomously. To get the users into the habit of setting goals, some systems send goal-setting reminders, potentially accompanied by new goals recommendations. Similarly, to recommend appropriate PA programs to the user, some systems inquire users about their goals and provide PA suggestions that make it possible to achieve these goals. These are examples of how a user can achieve a target behavior, e.g., set up a step goal or follow a PA recommendation, step-by-step with the system's help.

\textbf{Simulation} is implemented as cause-and-effect metaphors of growing gardens \cite{Consolvo2008d,Gouveia2016b,Hochsmann2019e} that bloom with PA and wither with inactivity, pet avatars \cite{King2013,King2016,Paul2017,Tong2017b} and human avatars \cite{Schafer2018,Michael2020} that are happy and thriving when the user is active and sad otherwise, and other virtual experiences \cite{Gremaud2018b,Tu2019b,fuemmeler2020mila,nicholas2021development} that are affected by real-world PA. Moreover, it is achieved through the presentation of expected health outcomes (short-term or long-term) based on current PA levels \cite{Gouveia2014,Rabbi2015d,Gouveia2015a,Cambo2017}, or via connecting locations or hours of the day with levels of PA or sedentariness \cite{VanDantzig2018c,Arrogi2019}. The concept behind the investigated papers that utilize simulation is that the user will develop an emotional bond and a sense of responsibility about the virtual garden or pet, which will drive them to perform more PA to better care for them.

Finally, \textbf{Rehearsal} is achieved through demonstration of exercises via short videos \cite{Klausen2016c,Alsaqer2017,Kooiman2018b} or animated icons \cite{Lee2016,Kim2018c}. This technique can be beneficial for novice and less tech-savvy target groups, such as the elderly, \rev{but advanced or sporty users might find it indifferent.}

\subsubsection{Dialogue}\label{d}\hfill\\
In the PSD Framework, Dialogue includes seven persuasive techniques: Rewards, Suggestion, Reminders, Similarity, Praise, and Social Role (See Table~\ref{tab:psd}).

In the included papers, \textbf{Rewards} are implemented as free game commodities \cite{Berkovsky2010a,Walsh2014a,Garde2015,Leinonen2017b,Morrison2017,Hochsmann2019e,Tu2019b,fuemmeler2020mila}, congratulatory feedback for goal achievement or breaking sedentariness bouts \cite{Mutsuddi2012c,King2013,AlAyubi2014,Bond2014,Zuckerman2014a,Ding2016,Poirier2016a,Walsh2016,Lee2016,Zhou2018c,Cauchard2019b,Liew2020b,larsen2020feasibility,kim2022effect}, additional system functionality \cite{King2013,King2016}, badges or points \cite{Chen2014b,Chen2016,Zuckerman2014a,Arigo2015,Blackman2015b,Katule2016b,Poirier2016a,Klausen2016c,Mendoza2017,Cambo2017,Patel2017a,Leinonen2017b,Zhao2017b,Tong2017b,Korinek2018,Mitchell2018b,Altmeyer2018b,Gremaud2018b,Corepal2019b,Ciravegna2019b,Arrogi2019,Nuijten2019,Galy2019b,Edney2019b,Tu2019b,Wernbacher2020,fuemmeler2020mila}, raffle tickets \cite{Washington2014a,Patel2016b,Patel2016,Brett2017,Kim2018c}, and material \cite{Patel2017a,Corepal2019b} or financial incentives \cite{Kramer2019,Finkelstein2015e,Harkins2017,Chung2017a,Chokshi2018b,Elliott2019b,Memon2018b,Corepal2019b,Mason2018b}. Based on our article pool, rewards is the most ambiguous PSD technique. Various large-scale studies with financial incentives have reported statistically non-significant improvements in the user's PA, while studies utilizing virtual rewards, such as points and badges have reported mixed results. Hence, this technique should be used with caution in future work and always in combination with others. \rev{ST technologies should promote a long-term usage beyond rewards, as corroborated by related work \cite{Rapp2014}. Nevertheless, gained rewards create stored value for the user of a ST system, increasing the need to stay engaged \cite{eyal2014hooked}. 
}

\textbf{Suggestion} takes the form of PA recommendations \cite{FesslerMichaelB.;RudelLawrenceL.;Brown2008a,Mutsuddi2012c,Bickmore2013,Pilloni2013,Gouveia2014,Kramer2019,Rabbi2015e,Rabbi2015d,Gouveia2015a,Agboola2016c,Klausen2016c,Boratto2017,Brett2017,Leinonen2017b,Zhao2017b,Chung2017a,Kramer2017b,Simons2018d, Pope2018b,VanDantzig2018c,Kooiman2018b,Pope2019b,Corepal2019b,Choi2019b,Klasnja2019,Galy2019b,VanBlarigan2019b,Middelweerd2020b,Wernbacher2020,fuemmeler2020mila,nicholas2021development}, exercise plans and guidance \cite{Lee2015b,Alsaqer2017,Robinson2019,Lee2016,nicholas2021development}, break and stretching suggestions \cite{VanDantzig2013a,Kim2013a,Bond2014,Finkelstein2015d,Cambo2017,Kim2018,Arrogi2019}, goal adjustment recommendation \cite{AlAyubi2014,Poirier2016a,Verbeek2012a,Bianchi-Hayes2018b,Middelweerd2020b,ek2020effectiveness}, behavior change tips \cite{King2013,ek2020effectiveness}, emergency services communication in case of injury \cite{park2021mobile}, or generally healthy living and self-care recommendations \cite{Hebden2014a,Kim2013a,stephens2022feasibility}. In the majority of the investigated articles, suggestions were limited to scripted tips and PA recommendations by the research team or PA experts. However, this one-size-fits-all approach is outdated in the era of ML and personalization, where suggestions can be micro-targeted \rev{and tailored to the needs of a specific user or group} to increase their effectiveness.

\textbf{Reminders} are implemented via automated phone calls \cite{FesslerMichaelB.;RudelLawrenceL.;Brown2008a}, text messages \cite{McNab2012b,VanDantzig2013a,Wang2015d,Agboola2016c,Mendoza2017,Chen2017,Polgreen2018b,Tong2019b,Gell2020a,VanBlarigan2019b}, e-mails \cite{Brett2017,Klasnja2019,Robinson2019}, social media notifications \cite{Chung2017a}, random or just-in-time notifications \cite{Gouveia2014,AlAyubi2014,AlAyubi2014,Gouveia2015a,Katule2016b,Cambo2017,Lyons2017b,Alsaqer2017,Zhou2018c,Ciravegna2019b,Klasnja2019,Zhang2019a,Edney2019b,nicholas2021development,koontz2021increasing}, watch reminders \cite{Lee2016,Kim2018c}, and visual, audio or haptic prompts and in-app reminders \cite{Bond2014,Pellegrini2015,Finkelstein2015d,Liu2016,Arrogi2019}. Their purpose includes reminding users of goal-setting, wearable wear time and activity logging, application usage instructions, sedentariness levels, break times, and current PA levels. However, reminders can be a double edged sword. If not sent parsimoniously, they can be ignored or cause annoyance to the user. To be effective, a reminder should be sent at a time when a user is ready to receive it. Such reminders fall under the umbrella of Just-in-Time Adaptive Interventions (JITAIs), a field that has gathered scientific attention \cite{hardeman2019systematic} and should be the focus of future work in the field of ST. 

\textbf{Similarity} is achieved through the use of embodied conversational agents \cite{Bickmore2013}, human avatars \cite{King2013,King2016,Alsaqer2017,Kim2018c,Galy2019b,Michael2020,fuemmeler2020mila}, or the utilization of the user's physical location \cite{Leinonen2017b}. \rev{Human-like representations, such as avatars, have the potential to provide a user experience that resembles human-to-human interaction, triggering social responses from the users. In other words, a human-controlled avatar yields social presence, namely the perception that another individual is in the user's environment \cite{lombard1997heart}. At the same time, the perceived ``human control'' that avatars seem to exert elicits stronger behavioral responses for users than the perception of machine control \cite{fox2015avatars}.}

\textbf{Praise} is achieved through motivational messages \cite{FesslerMichaelB.;RudelLawrenceL.;Brown2008a,Mutsuddi2012c,Bickmore2013,Kim2013a,Martin2015b,Ding2016,Agboola2016c,Mendoza2017,Alsaqer2017,Kim2018c,Mutsuddi2012c,Ciravegna2019b,nicholas2021development,damschroder2020effect,park2021mobile,koontz2021increasing}, feedback depending on goal achievement \cite{Patel2017a,Monroe2017b,Leinonen2017b,Kramer2017b,fuemmeler2020mila,kim2022effect}, or happy icons and emojis \cite{Wally2017,Paul2017}. Similarly to reminders, praise should not be too intrusive or repetitive, as it can become an annoyance to the user.

\textbf{Liking} takes the form of stylized, interactive displays \cite{Consolvo2008d,AlAyubi2014,Tong2017b}, customizable displays \cite{Zhao2017b}, enhanced usability \cite{Pilloni2013,Ciravegna2019b}, imaginary scenery interfaces \cite{Katule2016b,Gouveia2016b,Hochsmann2019e,Wernbacher2020,fuemmeler2020mila}, and user-tailored interface design \cite{Leinonen2017b,Mutsuddi2012c,Galy2019b}. Accessible and usable interfaces are of vital importance in the field of ST. Frequently, users access the applications while performing PA, e.g., at the gym or while running, which means that their design should be easy and intuitive. Complicated or confusing interfaces can be unappealing and may soon lose user interest. 

Finally, \textbf{Social Role} refers to utilizing the system as an accountability mechanism, such as a virtual or human coach or physician \cite{FesslerMichaelB.;RudelLawrenceL.;Brown2008a,VanDantzig2013a,Chen2017,Boratto2017,stephens2022feasibility,larsen2020feasibility}, a peer leader \cite{nicholas2021development}, or a virtual pet \cite{Blackman2015b,Katule2016b,King2016,Paul2017,Zhao2017b,Tong2017b}. \rev{Based on the model of supportive accountability \cite{mohr2011supportive}, accountability mechanisms, such as external human coaches, can foster motivation, encouragement, and ultimately adherence and UE.}

\subsubsection{Social Support}\label{ss}\hfill\\
In the PSD Framework, Social Support includes seven persuasive techniques: Social Comparison, Social Learning, Cooperation, Competition, Recognition, and Social Facilitation (See Table~\ref{tab:psd}).

\textbf{Social Comparison} takes the form of public performance displays \cite{Lim2011a,Altmeyer2018b,Esakia2018c}, virtual and real-world competitors \cite{Berkovsky2010a,Arigo2015,Leinonen2017b}, as well as performance sharing and comparison \cite{VanDantzig2013a,Khalil2013,King2013,Nishiyama2014,Chen2014b,Chen2016,AlAyubi2014,Wally2017,Babar2018b,Corepal2019b,Tong2019b,Nuijten2019,Zhang2019a,Middelweerd2020b,Foster2010a,Poirier2016a,nicholas2021development}. \rev{Users who are exposed to social comparison information desire to avoid the stigma of unhealthy behavior, such as decreased PA, and hence consider adapting their behavior to the majority rule \cite{yun2011social}.}

\textbf{Cooperation} is usually implemented through user team-ups (dyads or groups) \cite{King2013,Nishiyama2014,Chen2014b,Chen2016,Garde2015b,Blackman2015b,Katule2016b,King2016,Patel2016b,Patel2016,Brett2017,Leinonen2017b,Ren2018b,Esakia2018c,Corepal2019b,Cadmus-Bertram2019b,Choi2019b,Galy2019b,Liew2020b,fuemmeler2020mila,nicholas2021development}. The investigated articles reported better results when the groups consist of friends, colleagues or family members rather than random users. \rev{In other words, social connectedness is key for the success of cooperation-based interventions.}

\textbf{Social Learning} is achieved via PA-related discussion forums \cite{King2013,Hebden2014a,Arigo2015,King2016,Monroe2017b,Tong2019b,stephens2022feasibility} and social network groups \cite{Mendoza2017,Althoff2017b,Pope2018b,Babar2018b,Pope2019b,Corepal2019b,Edney2019b,Valentiner2017}, instant messaging functionality \cite{Chen2014b,Chen2016,AlAyubi2014,Garde2015b,Monroe2017b,Chung2017a,Tong2019b,Zhang2019a,Liew2020b,fuemmeler2020mila,stephens2022feasibility}, public profiles \cite{AlAyubi2014,Poirier2016a,Lyons2017b}, multi-player gaming mode \cite{Valentiner2017}, real-world support \cite{Katule2016b,Bianchi-Hayes2018b},  or peers' performance and experiences sharing \cite{Ren2018b,VanWoudenberg2020b}. By sharing their experiences, users can seek support, motivation and feel less alone in their HBC journey. However, some of the investigated papers report that users with social ties (e.g., colleagues, family members) did not utilize the social functionality as much, but preferred face-to-face communication instead. Hence, the this technique's importance may vary depending on the use case.

\textbf{Competition} normally takes the form of individual and group-based PA competitions and challenges \cite{Foster2010a,King2013,Nishiyama2014,Walsh2014a,Chen2014b,Chen2016,Zuckerman2014a,Garde2015b,Blackman2015b,King2016,Prestwich2017,Leinonen2017b,Shameli2019,Zhao2017b,Chung2017a,Esakia2018c,Gremaud2018b,Corepal2019b,Nuijten2019,Galy2019b,Edney2019b,Liew2020b,Michael2020,Wernbacher2020}. In our article pool, competitions are reported as more effective when they consist of users with similar PA levels. Competitions with large PA differences between participants can be deemed too easy for the more advanced users and unattainable by the less active. Similarly, when it comes to one-to-one competitions, the investigated papers report that they were more effective when the involved users had similar PA levels and limited step differences throughout the day. \rev{Hence, the competitors of the user can be either real or contrived to match the user's PA behavior.}

\textbf{Recognition} is implemented through competition leaderboards and podiums \cite{Foster2010a,Walsh2014a,Chen2016,Zuckerman2014a,Garde2015b,Arigo2015,Katule2016b,Prestwich2017,Brett2017,Shameli2019,Zhao2017b,Gremaud2018b,Corepal2019b,Edney2019b,Tu2019b,Liew2020b,Michael2020,Middelweerd2020b,stephens2022feasibility}, social network posts about winner teams or users \cite{Chung2017a}, physical awards \cite{Nuijten2019}, and success stories testimonies \cite{Brett2017,Alsaqer2017}. \rev{Evidently, Recognition is not a standalone intervention component but is usually combined with Competition or Social Comparison.}

\textbf{Social Facilitation} examples include social network commenting on PA-related posts \cite{Foster2010a,Poirier2016a,Mendoza2017,Lyons2017b,Babar2018b,Tu2019b,fuemmeler2020mila}, public testimonies \cite{Mutsuddi2012c}, virtual commodity swapping \cite{Morrison2017}, public participants' lists \cite{Gremaud2018b}, and user invitation schemes \cite{Leinonen2017b}. \rev{Note that Social Facilitation components can help increase UE by increasing user commitment. A user who performs a menial task, such as referring a friend to a fitness app, not only increases the user base, but in reality, invests in the product itself, creating stored value and promoting future use \cite{eyal2014hooked}.}

Finally, \textbf{Normative Influence} is achieved via public PA-related commitments \cite{Mutsuddi2012c,Patel2017a}, presentation of the financial and environmental effects of inactivity \cite{Mutsuddi2012c,Choi2019b}, virtual and physical users demonstrating ideal PA behavior \cite{King2013,VanWoudenberg2020b}, and comparison against PA guidelines \cite{Chen2014b,Chen2016,AlAyubi2014} and overall users' performance \cite{Liu2016,Patel2016b,Gouveia2016b,Wally2017,Brett2017,Leinonen2017b,Paul2017,yuasa2022estimation}. For example, based on the investigated literature, users think twice before they break PA commitment pledges to family and friends on social media, since this might negatively affect their social image. 

\subsubsection{System Credibility}\label{sc}\hfill\\
In the PSD Framework, Social Support includes seven persuasive techniques: Authority, Real-world Feel, Expertise, Trustworthiness, Surface Credibility, Third-party Endorsements, and Verifiability (See Table~\ref{tab:psd}). However, in the included papers, we only encounter the first four persuasive techniques.

In this review's article pool, \textbf{Authority} takes the form of external accountability mechanisms, such as human coaches \cite{FesslerMichaelB.;RudelLawrenceL.;Brown2008a,Pilloni2013} or physicians \cite{Martin2015b,Alsaqer2017,Kooiman2018b,stephens2022feasibility}, guideline recommendation by international organizations (e.g., WHO, US Health Agency) \cite{VanDantzig2013a,Lee2015b,Bianchi-Hayes2018b,yuasa2022estimation,park2021mobile}, as well as health organizations and committees behind the app creation \cite{Ciravegna2019b,Mason2018b}. Having high-profile organizations behind a system or functionality gives additional credibility to the system and enables the users to trust it more.

\textbf{Expertise} takes the form of PA-related content curated by domain experts \cite{Mutsuddi2012c,Alsaqer2017}, communication with human coaches and physicians \cite{Boratto2017,fuemmeler2020mila,larsen2020feasibility,damschroder2020effect}, or tech support portals \cite{Leinonen2017b}. Such functionality can be costly and thus is not commonly encountered in practice in scientific interventions.

\textbf{Trustworthiness} is implemented via regular updates \cite{Zhao2017b,Ciravegna2019b}, intensive testing and debugging \cite{Ciravegna2019b}, and data handling based on current regulations (e.g., GDPR) \cite{Wernbacher2020,park2021mobile}. \rev{While we rarely encounter this technique in the included works, system security is fundamental for the trusted use of ST technologies.}

\textbf{Real-word Feel} is achieved through counseling services with the researchers \cite{Lyons2017b,larsen2020feasibility,park2021mobile}, app store communication \cite{Ciravegna2019b}, or in-app and website contact forms \cite{Zhou2018c,Kooiman2018b,Galy2019b}.

\subsubsection{Others}\label{o}\hfill\\
Apart from PSD's persuasive techniques, we identify four additional techniques in the investigated papers: Goal-setting, Punishment, General Information, and Variability, which we explain further below.

\textbf{Goal-setting} is implemented as static or dynamic PA goals usually in terms of steps, active minutes or MVPA duration \cite{Consolvo2008d,Foster2010a,Bickmore2013,Pilloni2013,King2013,Gouveia2014,Kramer2019,Chen2014b,Chen2016,AlAyubi2014,Zuckerman2014a,Vathsangam2014,Arigo2015,Lee2015b,Gouveia2015a,Ding2016,Liu2016,Bronikowski2016,King2016,Poirier2016a,Walsh2016,Patel2016b,Patel2016,Lee2016,Verbeek2012a,Mendoza2017,Chen2017,Harkins2017,Boratto2017,Patel2017a,Monroe2017b,Brett2017,Lyons2017b,Leinonen2017b,Paul2017,Zhao2017b,Morrison2017,Chung2017a,Kramer2017b,Tong2017b,Korinek2018,Bianchi-Hayes2018b,Simons2018d,VanDantzig2018c,Zhou2018c,Mitchell2018b,Altmeyer2018b,Esakia2018c,Gremaud2018b,Chokshi2018b,Elliott2019b,Kooiman2018b,Polgreen2018b,Ciravegna2019b,Cadmus-Bertram2019b,Choi2019b,Hochsmann2019e,Galy2019b,Gell2020a,VanBlarigan2019b,Cauchard2019b,Robinson2019,Middelweerd2020b,Wernbacher2020,ek2020effectiveness,fuemmeler2020mila,nicholas2021development,stephens2022feasibility,larsen2020feasibility,wang2021smartphone,damschroder2020effect,koontz2021increasing}. This technique can overlap with the Personalization technique regarding personalized goals or with the Reduction technique regarding multiple, simultaneous goals. However, it is more generic, covering also a large number of papers which utilize static goals, which follow the international guidelines for PA. Static goals though apply the one-size-fits-all approach and cannot adjust to the user's changing needs and wants. Thus, future research should focus on tailoring and personalizing goals to the users' reality, including, but not limited to, habits, physical condition, personal expectations, daily schedule and location.

\textbf{Punishment} takes the form of negative visual or textual feedback for under-performance \cite{Kramer2017b,Kim2018c,Lee2016,Wally2017,Paul2017,yuasa2022estimation}, and virtual or monetary reward loss for goal accomplishment failure \cite{Chokshi2018b,Patel2017a}. While most of the investigated papers utilize positive feedback to promote HBC, it is unclear whether positive or negative feedback leads to more favorable BC in an HBC intervention. HBC theories make contradicting predictions regarding the influence of the feedback polarity \cite{kramer2017using}. In the PAST component, Praise has a higher PAST\_score than Punishment based on our article pool results, but either can potentially yield positive results.

\textbf{General Information} provision takes the form of e-mails or notifications that do not necessarily reflect on the user's performance, rather than provide general information and interesting facts regarding PA and HBC \cite{Simons2018d,Leinonen2017b,Paul2017,Chung2017a,Kooiman2018b,Arrogi2019,VanBlarigan2019b,ek2020effectiveness,fuemmeler2020mila,yuasa2022estimation,nicholas2021development,stephens2022feasibility,damschroder2020effect,kim2022effect}.

Finally, \textbf{Variability} refers to the system's ability to provide a variable experience to the user through variable rewards \cite{Morrison2017,Leinonen2017b,Paul2017}, variable game elements, e.g., levels, varying interfaces, and hidden tasks \cite{Leinonen2017b,fuemmeler2020mila,nicholas2021development}. Practitioners have praised the power of variability in sustaining UE \cite{eyal2014hooked}, but it is still not fully utilized in ST for HBC research. The limited related papers report neutral results, but future research should focus more on identifying the effect of variability on HBC and UE. 

\subsubsection{PAST: A Periodic Table of Successful Health Behavior Change}\label{periodicTable}\hfill\\
Designing ST products is far from straightforward, which is proven by current ST technology's pitfalls (e.g., dubious effectiveness and high attrition rates) as discussed in Section~\ref{introduction}, and the variety of persuasive techniques utilized in practice as demonstrated in Sections \ref{pts} to \ref{o}. The PSD Framework (Section~\ref{hbc}), which we use to analyze the persuasive techniques applied in the included papers, aims to aid practitioners in this intricate design process by presenting 28 design principles for persuasive system content and functionality (Table~\ref{tab:psd}). However, PSD attaches the same weight to all 28 principles, ranging from self-monitoring to third-party endorsements. It is evident, though, that some principles (e.g., self-monitoring, reminders, personalization) may bear higher importance than others in ST technology. Here is where the PAST component comes in. 
\begin{table}[htb!]
\resizebox{0.7\textwidth}{!}{%
\begin{tabular}{|l|l|}
\hline
\rowcolor[HTML]{524C99} 
\multicolumn{1}{|c|}{\cellcolor[HTML]{524C99}{\color[HTML]{FFFFFF} \textbf{Notation}}} & \multicolumn{1}{c|}{\cellcolor[HTML]{524C99}{\color[HTML]{FFFFFF} \textbf{Explanation}}} \\ \hline
$t_i$                                                                                  & Technique $i$, $t_i\in\{1,2,\ldots,32\}$                                                                            \\ \hline
$n_{papers}$                                                                                  & Total number of papers                                                                                \\ \hline
$p_{i,j}$                                                                              & Paper $j$ with technique $i$, 1 if technique $i$ appears in paper $j$, 0 otherwise                                                               \\ \hline
$r_{i,j}$                                                                                & Result of paper $j$ with technique $i$, $r_{i,j}\in\{-1, -0.5, 0, 0.5, 1\}$                                                                      \\ \hline
$w$                                                                                
& The preferred weight for the PAST\_score, $w\in[0,1]$                                                                       \\ \hline
\end{tabular}%
}\caption{Notation for the score generation functions of the PAST component.\label{tab:notations}}
\vspace{-6mm}
\end{table}

\noindent\textbf{Not all techniques are equally important:} All design techniques in ST systems can be significant under different settings, and no single technique can guarantee success. However, some techniques are more efficient than others or may have specific characteristics that make them more or less frequently used in practice. To capture the importance of ST design techniques, we devise an evidence-based score for each technique based upon the investigated papers' results. 

Table~\ref{tab:notations} provides the necessary notation for the understanding of the score generation formulas. We define a value $r_{i,j}$ for the result of an investigated paper $j$ that uses a technique $i$ as follows: $-1$ corresponds to a negative result, $-0.5$ a partially negative result, $0$ a neutral result, $0.5$ a partially positive result, and $1$ a positive result. A result is considered positive if the respective paper's intervention led to a statistically significant increase in the users' PA. It is considered neutral if the intervention caused no statistical change in the users' PA and negative if it led to a statistically significant decrease in the users' PA. Finally, we consider a result partially positive or negative if the paper reported mixed results. We manually perform the numerical coding of the results. Note that if a paper $j$ reports positive results ($r_{i,j}=1$) and utilizes two different techniques $t_1$ and $t_2$, then the positive result will be split between the two techniques, assuming all techniques have equal contribution to the final result; namely, $r_{1,j}=0.5$ and $r_{2,j}=0.5$. 

Then, we define the efficacy of the technique $t_i$, as the sum of the coded results of the papers the technique appears in divided by the number of these papers, and the frequency of the technique $t_i$, as the number of papers that the technique appears divided by the total number of papers:
\begin{equation}
\textrm{efficacy}(t_i)=\frac{\sum_j r_{i,j}}{\sum_{j} p_{i,j}}
\quad\text{and}\quad 
\textrm{frequency}(t_i)=\frac{\sum_{j} p_{i,j}}{\sum_{j} p_{j}}
\end{equation}
The final score for a technique is a combination of its reported efficacy and usage frequency in the papers: 
$$\textrm{PAST\_score}(t_i)=w*\textrm{efficacy}(t_i)+(1-w)*\textrm{frequency}(t_i)$$
where $w$ is a weight that can be used to emphasize the efficacy or the frequency metric. We deemed this weighted combination necessary to avoid over-rewarding infrequent techniques for individual positive results or over-rewarding frequent techniques only for their prevalence in the investigated papers. In the remainder, for presentation purposes, we set $w=0.5$ to achieve a balance between the measured quantities and normalize the PAST\_score in the $[0,1]$ interval. 

\begin{figure}[htb!]
  \centering
  \includegraphics[width=.8\linewidth]{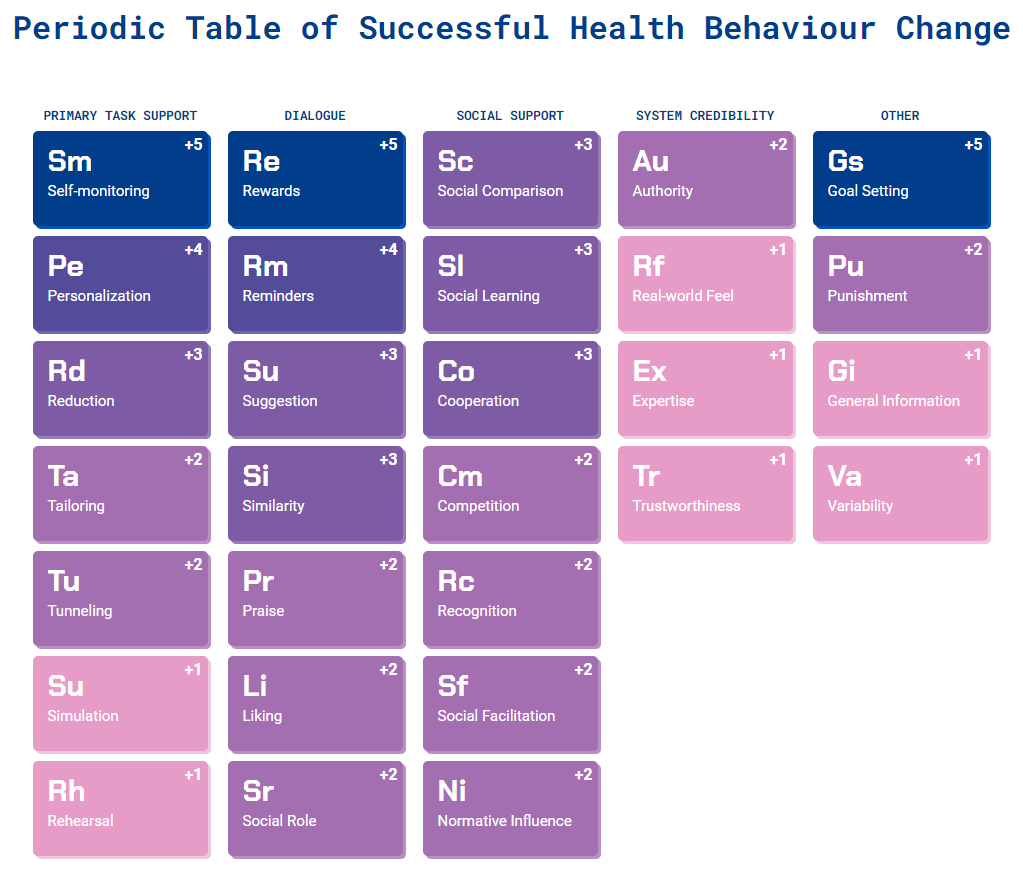}
  \caption{The Periodic Table of Successful Health Behavior Change (PAST). Each element is color-coded based on its assigned PAST\_score, which is visible on the top right corner.}
  \Description{The Periodic Table of Successful Health Behavior Change (PAST). Elements in the table are divided into four categories (Primary Task Support, Dialogue, Social Support, System Credibility) as per PSD. Each cell contains a technique's abbreviation, name, and PAST\_score.}
  \label{fig:periodictable}
\end{figure}

\noindent\textbf{The Periodic Table of Successful HBC elements:} We calculate the PAST\_scores for the techniques in the studied papers and create the Periodic Table of Successful Health Behavior Change, as seen in Figure~\ref{fig:periodictable}. Presenting these elements in a Periodic Table is human-friendly and easily memorizable. At the same time, it offers a comprehensive and formal classification, describing a user interface and system features design process. 

Specifically, each element in the table represents a PSD technique (Table~\ref{tab:psd}) or additional persuasive techniques encountered in the investigated papers, namely, goal-setting, punishment, general information provision, and variability (32 elements in total); as discussed in detail in Sections \ref{pts} to \ref{o}. Elements in our periodic table are divided into five categories (Primary Task Support, Dialogue, Social Support, System Credibility, Other), and are organized in columns, respectively. Each cell contains an abbreviation for the technique, the technique's name, as well as the technique's PAST\_score. Finally, we normalize the scores in a $[+1,+5]$ interval for the periodic table as follows: scores $\in[0,0.2)\rightarrow+1$, scores $\in[0.2,0.4)\rightarrow+2$, scores $\in[0.4,0.6)\rightarrow+3$, scores $\in[0.6,0.8)\rightarrow+4$, and scores $\in[0.8,1)\rightarrow+5$. Each technique is color-coded based on its normalized PAST\_score, and the PAST\_score is presented at the top-right corner of each cell. Similar to the Periodic Table of Elements, PAST aims to provide a comprehensive, evidence-based, and robust standard for designing HBC interventions. 
Our PAST\_score aims at guiding creators in the design process of ST technology, by helping them prioritize the development of certain ST features and functionalities (based on their expected impact on HBC and UE).

\noindent\textbf{Analysis and Insights:} Based on our periodic table, we analyze the reviewed techniques' scores, obtaining insights on their importance. Note that the score ranges between +1 and +5, with higher scores holding more persuasive power for HBC, e.g., a technique with a score of +5 can be seen as more important than one with a +1. Nevertheless, typically a combination of different techniques should be used in practice for an efficient system. First, in the periodic table of Figure~\ref{fig:periodictable}, we observe that not all categories are as frequent; Primary Task Support is the most frequent (91\% of investigated articles), followed by Dialogue (80\%), Social Support (42\%), and System Credibility (9\%). The disproportionate frequencies already indicate the relative importance of certain feature categories (e.g., Primary Task Support) over others serving a complementary role (e.g., System Credibility).

Within the Primary Task Support category, Self-monitoring is the most effective technique (+5), followed by Personalization (+4), Reduction (+3), Tailoring, and Tunneling (+2), and finally, Simulation and Rehearsal (+1). In the Dialogue category, Rewards are the most effective persuasive technique (+5), followed by Reminders (+4), Suggestion and Similarity (+3), and finally, Praise, Liking, and Social Role (+2). In the Social Support category, Social Comparison, Social Learning, and Cooperation share similar weight (+3), followed by Competition, Recognition, and Social Facilitation and Normative Influence (+2). In the System Credibility Category, Authority is the most effective technique (+2) (even though it still holds limited persuasive power), followed by Real-world Feel, Expertise, and Trustworthiness (+1). In contrast, the Surface Credibility, Third-party Endorsements, and Verifiability techniques are not explicitly mentioned in any investigated articles. Note that the very low frequency of these techniques in the investigated papers and the lack of research on the persuasive power of system credibility features might be responsible for the low importance of this category's techniques. Finally, among the techniques that do not fall under the PSD framework, Goal-setting is the most effective (+5), followed by Punishment (+2), General Information provision (+1), and Variability (+1).

For presentation purposes, in our discussion, we use a weight $w=0.5$ for the PAST\_score. However, the PAST\_score weights can be adjusted according to application needs, emphasizing the efficacy or frequency aspects. To enable researchers and practitioners to adjust the PAST component according to their application requirements, we offer an online, interactive exploration tool for weight adjustment, \rev{filtering and real-time visualization of the adapted PAST\_scores}, where the user can explore how different weights affect the positioning of persuasive techniques in the PAST table. 

\subsection{Evaluating User Engagement \& the SELF Component}\label{self}
The PAST Component helps practitioners and researchers design evidence-based ST technology to maximize its effect on user HBC. To verify this contribution, we need a way to measure user experience, beyond simple measurement of PA changes \cite{rooksby2014personal}. Notice that the user experience incorporates different facets of the user's behavior, such as the user's thoughts and emotions, their PA and behavioral patterns, and the user's interaction with the system and the environmental factors that affect it. However, as discussed in Section~\ref{introduction}, there is no standard approach for evaluating the effects of ST interventions on user habits and UE. Moreover, as seen in Section~\ref{ue}, current UE metrics focus on web or sometimes mobile applications rather than wearable devices. Also, existing metrics focus on quantifying the users' purchasing or browsing behavior rather than their health behavior. This is where the SELF component of our PAST SELF framework comes in.

To find out which measures and metrics can be utilized to evaluate ST technology, we identified and extracted all measured quantities mentioned in the investigated papers (e.g., self-reported data, PA data, UE data, metadata) and categorized them based on four UE aspects (inspired by the work of \citet{lalmas2014measuring} discussed in Section~\ref{ue}): \rev{the Perceived Self Aspect, the Physical Self Aspect, the Behavioral Self Aspect, and the Environmental Aspect. The \textbf{Perceived Self Aspect} (See SELF Table~\ref{tab:emotional}) refers to the user's self-reported image of their life regarding their everyday experiences, as well as psychological, technological, social, and health factors, and is usually measured through qualitative evaluation methods.} Hence, in our work, it encompasses all self-reported data utilized in our article pool. The \textbf{Physical Self Aspect} (See SELF Table~\ref{tab:cognitive}) refers to the user's physical reaction to the interaction with system, which in ST technology can be for example interpreted as the PA performed in response to the system's intervention. Thus, it includes all PA-related quantities from the investigated articles. The \textbf{Behavioral Self Aspect} (See SELF Table~\ref{tab:behavioral}) refers to the user's behavioral response to the system, which in ST takes the form of UE metrics, such as wear-time, and session duration. Finally, the \textbf{Environmental Aspect} (See SELF Table~\ref{tab:environmental}) refers to external factors that may affect the user's interaction with the system, such as weather or location. Hence, we identified and incorporated in this aspect all metadata reported in the investigated articles. \rev{Note that these aspects draw from and adapt Lalmas' work to the realities of ubiquitous computing and ST technology, and they expand it to account for the particularity of the domain. For example, we introduce a novel aspect of ubiquitous UE, namely the ``Environmental Aspect'', to incorporate various external factors that influence a user's behavior and interaction with the ST system, such as the weather or their daily schedule. These factors might not be central to a user's interaction with a traditional computer system. However, they are detrimental to the user's PA behavior and, subsequently, their interaction with ST technology.}

This categorization has led to the creation of the \textbf{S}elf-Tracking \textbf{E}va\textbf{l}uation \textbf{F}ramework (SELF). SELF is a novel tool for the standardization of the evaluation process of ubiquitous ST technology-based interventions. Each aspect of the SELF component is shown here through a SELF Table. Each SELF Table contains the measures and metrics related to its respective aspect. Note that the SELF Tables are not supposed to be exhaustive lists of all potential ST metrics. On the contrary, they constitute clear and organized presentations of the most commonly used metrics in the ST literature, as identified during our review process. 

\subsubsection{The \rev{Perceived Self} Aspect of UE}\hfill\\
SELF Table~\ref{tab:emotional} presents the measurable concepts related to the \textbf{Perceived Self Aspect of UE}. ST and HBC research is highly interdisciplinary and almost always utilizes self-reports. We identified that these reports are centered around five main factors: User Factors, Psychological Factors, Human-Computer Interaction Factors, Social Factors, and Health Factors. Each Factor consists of the following measurable concepts:  
\begin{enumerate}
    \item \textbf{\textit{User factors}} include concepts such as technological habits \& competency \cite{Consolvo2008d,Berkovsky2010a,Mutsuddi2012c,McNab2012b,Vathsangam2014,Garde2015,Arigo2015,Rabbi2015e,Wang2015d,Ding2016,Verbeek2012a,Chen2017,Zhao2017b,Chung2017a,Tong2017b,Kim2018c,Tong2019b,Gell2020a,Wernbacher2020}, PA habits \& competency \cite{Consolvo2008d,FesslerMichaelB.;RudelLawrenceL.;Brown2008a,Mutsuddi2012c,Khalil2013,Wu2013,King2013,Washington2014a,Walsh2014a,Kramer2019,Hebden2014a,Kim2013a,AlAyubi2014,Finkelstein2015e,Finkelstein2015d,Martin2015b,Wang2015d,Bronikowski2016,King2016,Klausen2016c,Patel2016,Wally2017,Prestwich2017,Brett2017,Alsaqer2017,Zhao2017b,Tong2017b,Korinek2018,Bianchi-Hayes2018b,Simons2018d,Pope2018b,Zhou2018c,Altmeyer2018b,Elliott2019b,Kooiman2018b,Kim2018c,Choi2019b,Hochsmann2019e,Robinson2019,Edney2019b,Michael2020}, sedentary habits \cite{King2013,Hebden2014a,King2016,Pope2018b}, dietary habits \cite{Chung2017a,Pope2019b}, daily life patterns \cite{Tong2017b,Korinek2018,Bianchi-Hayes2018b,Yoon2018b,Arrogi2019,Cauchard2019b,Wernbacher2020}, time perspective \cite{Harkins2017,Corepal2019b}, and habit formation \cite{Korinek2018,Tu2019b}. Such information can be utilized to contextualize the system to the user's reality, for example, by providing personalized suggestions and JITAIs based on the user's schedule and daily habits.
    \item \textbf{\textit{Psychological factors}} include concepts, such as personality traits \cite{Lim2011a,Korinek2018,Nuijten2019,VanDantzig2013a}, self-efficacy \cite{FesslerMichaelB.;RudelLawrenceL.;Brown2008a,McNab2012b,Simons2018d,Kooiman2018b,Schafer2018,Memon2018b,Pope2019b,Choi2019b,Robinson2019,VanWoudenberg2020b,Middelweerd2020b}, the user's stage of behavior change \cite{Consolvo2008d,Lim2011a,Mutsuddi2012c,McNab2012b,Hebden2014a,Rabbi2015d,Gouveia2015a,Ding2016,Agboola2016c,Gouveia2016b,Tong2017b,Bianchi-Hayes2018b,Ren2018b,Memon2018b,Choi2019b}, behavioral self-regulation \cite{Mendoza2017,Prestwich2017,Schafer2018,Middelweerd2020b}, motivation \cite{Katule2016b,Brett2017,Lyons2017b,Pope2019b,Hochsmann2019e,Michael2020,VanWoudenberg2020b}, emotional state \cite{Rabbi2015d,Ding2016,Robinson2019}, user goals \& expectations \cite{FesslerMichaelB.;RudelLawrenceL.;Brown2008a,Nishiyama2014,Rabbi2015d,Korinek2018,Bianchi-Hayes2018b,Middelweerd2020b}, attitudes \& intentions toward HBC \cite{Khalil2013,Pilloni2013,King2013,Walsh2014a,Arigo2015,Rabbi2015d,Wang2015d,Mendoza2017,Harkins2017,Prestwich2017,Brett2017,Lyons2017b,Korinek2018,Korinek2018,Bianchi-Hayes2018b,Simons2018d,Pope2018b,Zhou2018c,Schafer2018,Pope2019b,Choi2019b,Arrogi2019,Galy2019b,Cauchard2019b,Robinson2019,Middelweerd2020b}, attitudes toward technology \cite{Wang2015d,Lyons2017b}, as well as attitudes toward one's appearance \cite{Memon2018b}. This factor is vital in understanding the user behavior regarding ST, as human psychology influences human behavior. For instance, applied research has reported "dramatic improvements in recruitment, retention, and progress using [BC] stage-matched interventions" \cite{prochaska1997transtheoretical}. 
    \item \textbf{\textit{HCI factors}} include system usability \cite{VanDantzig2013a,Pilloni2013,Nishiyama2014,AlAyubi2014,Zuckerman2014a,Vathsangam2014,Cambo2017,Alsaqer2017,Simons2018d,Altmeyer2018b,Kim2018c,Tong2019b,Arrogi2019,Middelweerd2020b,Wernbacher2020}, system utility \cite{Vathsangam2014,Pellegrini2015,Bianchi-Hayes2018b}, and user expectations \& satisfaction \cite{Berkovsky2010a,King2013,Gouveia2014,Zuckerman2014a,Pellegrini2015,Rabbi2015e,Lee2015b,Martin2015b,King2016,Lee2016,Pope2018b,Esakia2018c,Gremaud2018b,Kooiman2018b,Schafer2018,Pope2019b,Cadmus-Bertram2019b,Nuijten2019,Gell2020a,VanBlarigan2019b,VanWoudenberg2020b,Middelweerd2020b}. Such measures enable researchers to evaluate the effect of different features and UI components on HBC and UE, e.g., through A/B Testing.
    \item \textbf{\textit{Social factors}} measure concepts, such as social support \cite{FesslerMichaelB.;RudelLawrenceL.;Brown2008a,Bronikowski2016,Harkins2017,Bianchi-Hayes2018b,Simons2018d,Esakia2018c,Pope2019b,Corepal2019b,Choi2019b}, social influence \& norms \cite{Khalil2013,Kooiman2018b,Corepal2019b,Middelweerd2020b}, social comparison tendencies \cite{Arigo2015}, and group cohesion \& closeness \cite{Bianchi-Hayes2018b,Esakia2018c,VanWoudenberg2020b}. \rev{While the focus of ST technology has been on individual HBC, research supports that ST is a profoundly social practice \cite{Lupton2014,rooksby2014personal}, and hence social factors play an important role on a user's HBC journey.} 
    \item \textbf{\textit{Health factors}} incorporate concepts such as physical health \cite{Washington2014a,Finkelstein2015d,Chen2017,Harkins2017,Monroe2017b,Pope2018b,Zhou2018c,Kooiman2018b,Cadmus-Bertram2019b,Mendoza2017,Robinson2019}, mental health \cite{Bickmore2013,Agboola2016c,Chen2017,Alsaqer2017,Corepal2019b}, health literacy \cite{Bickmore2013}, and quality of life \cite{Finkelstein2015e,Mendoza2017,Chen2017,Leinonen2017b,Korinek2018,Pope2018b,Chokshi2018b}. Note that people with different health issues might require PA programs tailored to their condition. Hence, acquiring health-related information might be crucial for providing safe and effective PA recommendations through ST technology.
\end{enumerate}
\arrayrulecolor{white}
\setlength\arrayrulewidth{3pt}
\def\arraystretch{1.2}
\begin{table}[htb!]
\resizebox{.65\textwidth}{!}{%
\begin{tabular}{|Sl|Sl|}
\hline
\rowcolor[HTML]{3C3E69} 
\multicolumn{2}{|c|}{\cellcolor[HTML]{3C3E69}{\color[HTML]{FFFFFF} \textbf{PERCEIVED SELF ASPECT OF UE}}} \\ \hline
\rowcolor[HTML]{A1A3E0} 
{\color[HTML]{FFFFFF} \textbf{User Factors}} & {\color[HTML]{FFFFFF} \textbf{Psychological Factors}} \\ \hline
\rowcolor[HTML]{F2F2FB} 
\begin{tabular}[c]{@{}l@{}}\tabitem Demographics\\ \tabitem Technology Habits \& Competency\\ \tabitem PA Habits \& Competency\\ \tabitem Sedentary Habits\\ \tabitem Dietary Habits\\ \tabitem Daily Life\\ \tabitem Time Perspective\\ \tabitem Habit Formation\end{tabular} &

\begin{tabular}[c]{@{}l@{}}\tabitem Personality Traits\\ \tabitem Self-Efficacy\\ \tabitem Stage of Behavior Change\\ \tabitem Behavioral Self-Regulation\\ \tabitem Motivation\\ \tabitem Emotional State\\ \tabitem Goals \& Expectations\\ \tabitem Attitude \& Intentions toward HBC\\ \tabitem Attitudes toward Technology\\ \tabitem Attitudes toward Appearance \& Self-Esteem\end{tabular} \\ 

\rowcolor[HTML]{A1A3E0} 
{\color[HTML]{FFFFFF} \textbf{HCI Factors}} & \cellcolor[HTML]{F2F2FB} \\ \cline{1-1}
\rowcolor[HTML]{F2F2FB} 
{\color[HTML]{000000} \begin{tabular}[c]{@{}l@{}}\tabitem System Usability\\ \tabitem System Usefulness\\ \tabitem User Experience \& User Satisfaction\end{tabular}} & \multirow{-2}{*}{\cellcolor[HTML]{F2F2FB}} \\ 
\rowcolor[HTML]{A1A3E0} 
{\color[HTML]{FFFFFF} \textbf{Social Factors}} & {\color[HTML]{FFFFFF} \textbf{Health Factors}} \\ 
\rowcolor[HTML]{F2F2FB} 
\begin{tabular}[c]{@{}l@{}}\tabitem Social Support\\ \tabitem Social Influence \& Social Norm\\ \tabitem Social Comparison Tendencies\\ \tabitem Group Cohesion \& Closeness\end{tabular} & \begin{tabular}[c]{@{}l@{}}\tabitem Physical Health\\ \tabitem Mental Health\\ \tabitem Health Literacy\\ \tabitem Quality of Life\end{tabular} \\ \hline
\end{tabular}%
}
\caption{The \rev{Perceived Self} Aspect of UE: Self-reported data related to user factors, psychological factors, social factors, HCI factors, and health factors that have been utilized in ST technology for HBC and UE.\label{tab:emotional}}
\vspace{-6mm}
\end{table}

\rev{Self-reports can provide rich and meaningful information through user feedback, opening up new perspectives on user behavior. Qualitative data can reveal subtle yet critical design pitfalls, which, if discovered promptly, can help researchers improve the effectiveness of a system during user interventions. Hence qualitative studies are critical in the early stages of development of ST technology, and user interventions can be utilized once the system is mature enough to demonstrate effectiveness in the real-world \cite{Klasnja2011}.}

\rev{Upon handling qualitative data, though, one has to accept that the results may be intertwined with the experience of the researchers and that the findings might not generalize as well as large-scale quantitative studies \cite{Gerling2020}.} Additionally, a limitation of self-reports is that they rely on the user \rev{input.} However, grabbing the users' attention and requesting their time can be challenging in an era of attention scarcity \rev{\cite{franck2019economy}}. Thus, future work could focus on automating the extraction of knowledge related to \rev{some} factors of the \rev{Perceived Self} Aspect of UE. For instance, could an ML model predict the user's motivation or their attitude toward PA? 

\subsubsection{The \rev{Physical Self} Aspect of UE}\hfill\\
SELF Table~\ref{tab:cognitive} presents the metrics related to the \textbf{Physical Self Aspect of UE}, with a focus on PA. Hence, in the investigated papers, we identified PA-related measures capturing generic PA characteristics (e.g., Moderate to Vigorous Physical Activity (MVPA) duration, energy expenditure, heart rate) \cite{Consolvo2008d,Lim2011a,Berkovsky2010a,VanDantzig2013a,Wu2013,Pilloni2013,Washington2014a,Martin2015b,Rabbi2015d,Blackman2015b,Ding2016,Walsh2016,Mendoza2017,Cambo2017,Lyons2017b,Althoff2017b,Chung2017a,Bianchi-Hayes2018b,Pope2018b,Gremaud2018b,Pope2019b,Ciravegna2019b,Choi2019b,Hochsmann2019e,Klasnja2019,Arrogi2019,Nuijten2019,Michael2020,Wernbacher2020,Nishiyama2014,Hebden2014a,Bond2014,Pellegrini2015,Arigo2015,Rabbi2015e,Finkelstein2015e,Wang2015d,King2016,Klausen2016c,Lee2016,Boratto2017,Leinonen2017b,Alsaqer2017,Zhao2017b,Simons2018d,Pope2018b,Kooiman2018b,Kim2018c,Schafer2018,Yoon2018b,Corepal2019b,Cadmus-Bertram2019b,Hochsmann2019e,Galy2019b,Zhang2019a,Gell2020a,VanBlarigan2019b,Robinson2019,Edney2019b,Middelweerd2020b}, walking-specific characteristics (e.g., step count, floor count) \cite{Lim2011a,Foster2010a,FesslerMichaelB.;RudelLawrenceL.;Brown2008a,Mutsuddi2012c,McNab2012b,Bickmore2013,Khalil2013,Wu2013,King2013,Gouveia2014,Washington2014a,Walsh2014a,Kramer2019,Chen2014b,Chen2016,Hebden2014a,Kim2013a,AlAyubi2014,Garde2015,Arigo2015,Rabbi2015e,Lee2015b,Finkelstein2015d,Martin2015b,Rabbi2015d,Wang2015d,Blackman2015b,Katule2016b,Ding2016,Bronikowski2016,Agboola2016c,Poirier2016a,Walsh2016,Patel2016b,Gouveia2016b,Patel2016,Verbeek2012a,Mendoza2017,Chen2017,Wally2017,Harkins2017,Cambo2017,Prestwich2017,Patel2017a,Monroe2017b,Brett2017,Lyons2017b,Shameli2019,Paul2017,Althoff2017b,Morrison2017,Chung2017a,Kramer2017b,Tong2017b,Korinek2018,Bianchi-Hayes2018b,Simons2018d,Pope2018b,VanDantzig2018c,Zhou2018c,Mitchell2018b,Ren2018b,Altmeyer2018b,Esakia2018c,Gremaud2018b,Chokshi2018b,Elliott2019b,Kooiman2018b,Polgreen2018b,Memon2018b,Pope2019b,Yoon2018b,Corepal2019b,Ciravegna2019b,Mason2018b,Cadmus-Bertram2019b,Choi2019b,Hochsmann2019e,Tong2019b,Klasnja2019,Arrogi2019,Galy2019b,Zhang2019a,VanBlarigan2019b,Cauchard2019b,Robinson2019,Tu2019b,Liew2020b,VanWoudenberg2020b,Middelweerd2020b,Wernbacher2020}, and running-specific characteristics (e.g., running velocity, lap time) \cite{Rabbi2015e,Babar2018b}. In addition, we identified measurable concepts related to sedentariness (e.g., sedentary bout duration, number of sit-to-stand transitions) \cite{Hebden2014a,Bond2014,Pellegrini2015,Rabbi2015e,Finkelstein2015e,Arrogi2019,Gell2020a} or goal accomplishment (e.g., frequency of goal compliance) \cite{Zuckerman2014a,Liu2016,Patel2016b,Patel2016,Verbeek2012a,Harkins2017,Patel2017a,Bianchi-Hayes2018b,Zhou2018c,Ren2018b,Altmeyer2018b,Zhang2019a,Robinson2019}. 
\begin{table}[htb!]
\resizebox{.65\textwidth}{!}{%
\begin{tabular}{|Sl|Sl|}
\hline
\rowcolor[HTML]{3C3E69} 
\multicolumn{2}{|c|}{\cellcolor[HTML]{3C3E69}{\color[HTML]{FFFFFF} \textbf{PHYSICAL SELF ASPECT OF UE}}} \\ \hline
\rowcolor[HTML]{787CD2} 
\multicolumn{1}{|c|}{\cellcolor[HTML]{787CD2}{\color[HTML]{FFFFFF} \textbf{PHYSICAL ACTIVITY}}} & \multicolumn{1}{c|}{\cellcolor[HTML]{787CD2}{\color[HTML]{FFFFFF} \textbf{OTHERS}}} \\ \hline
\rowcolor[HTML]{A1A3E0} 
{\color[HTML]{FFFFFF} \textbf{Generic Activity Characteristics}} & {\color[HTML]{FFFFFF} \textbf{Sedentariness Characterists}} \\ 
\rowcolor[HTML]{F2F2FB} 
\begin{tabular}[c]{@{}l@{}}\tabitem Energy Expenditure (MET)\\ \tabitem Energy Expenditure (Calories)\\ \tabitem Heart Rate\\ \tabitem Number of Planned Activity Sessions\\ \tabitem Number of Executed Activity Sessions\\ \tabitem Time in-between Activity Sessions\\ \tabitem Activity Session Duration\\ \tabitem Total Activity Duration\\ \tabitem Number of Active Days\\ \tabitem Activity Intensity\\ \tabitem Activity Type\\ \tabitem MVPA Duration\\ \tabitem Number of MVPA Bouts\\ \tabitem MVPA Bout Duration\\ \tabitem Performance Improvement\end{tabular} & \begin{tabular}[c]{@{}l@{}}\tabitem Total Sedentariness Duration\\ \tabitem Sedentary Bout Duration\\ \tabitem Number of Sedentary Bouts\\ \tabitem Number of Sit-to-stand Transitions\end{tabular} \\ 
\rowcolor[HTML]{A1A3E0} 
{\color[HTML]{FFFFFF} \textbf{Walking-specific Characteristics}} & \multicolumn{1}{c|}{\cellcolor[HTML]{A1A3E0}{\color[HTML]{FFFFFF} \textbf{Goal Accomplishment Characteristics}}} \\ 
\rowcolor[HTML]{F2F2FB} 
\begin{tabular}[c]{@{}l@{}}\tabitem Step Count\\ \tabitem Floor Count\\ \tabitem Walking Duration\\ \tabitem Distance Walked\\ \tabitem Average Mile Time\\ \tabitem Walking Velocity\\ \tabitem Time in-between Walks\end{tabular} & {\color[HTML]{000000} \begin{tabular}[c]{@{}l@{}}\tabitem Goal Accomplishment Rate\\ \tabitem Frequency of Goal Compliance\end{tabular}} \\ 
\rowcolor[HTML]{A1A3E0} 
{\color[HTML]{FFFFFF} \textbf{Running-specific Characteristics}} & \cellcolor[HTML]{F2F2FB} \\ \cline{1-1}
\rowcolor[HTML]{F2F2FB} 
\begin{tabular}[c]{@{}l@{}}\tabitem Running Duration\\ \tabitem Running Distance\\ \tabitem Running Velocity\\ \tabitem Lap Time\end{tabular} & \multirow{-2}{*}{\cellcolor[HTML]{F2F2FB}} \\ \hline
\end{tabular}%
}
\caption{The \rev{Physical Self} Aspect of User Engagement: Physical activity, sedentariness and goal accomplishment in response to the user's interaction with ST technology.\label{tab:cognitive}}
\vspace{-6mm}
\end{table}

\subsubsection{The \rev{Behavioral Self} Aspect of UE}\hfill\\
SELF Table~\ref{tab:behavioral} presents the metrics related to the \textbf{\rev{Behavioral Self} Aspect of UE}. Usage analytics are commonly utilized to quantify users' behavior concerning the system. Usage metrics can be split into Intra-session metrics and Inter-session metrics. Inter-session metrics are more common in our article pool, and they usually take the form of wear time for activity trackers or total usage time for apps. However, both intra-session and inter-session metrics give us valuable information related to UE. Currently, the majority of the investigated papers do not incorporate this aspect of UE. However, future research should focus more on exploring such metrics' effect on the user's HBC journey. Intra-session metrics measure the user's UE with the ST technology during an individual session. A session is a set of user interactions with a system that takes place within a given time frame. Optimizations can take place based on intra-session metrics, with increasing complexity from "Feature" granularity (e.g., number of accesses of the individual PA goal-setting functionality) to "Session" granularity (e.g., number of PA goals set after a system reminder). Intra-session metrics can be categorized into three groups, Involvement metrics (e.g., session duration, screen views) \cite{Gouveia2014,Gouveia2015a,Gouveia2016b,Leinonen2017b,Zhao2017b,Simons2018d,Hochsmann2019e,Tong2019b,Cauchard2019b}, Interaction metrics (e.g., number of accepted notifications, notification response time) \cite{VanDantzig2013a,Hebden2014a,Gouveia2015a,AlAyubi2014,Pellegrini2015,Rabbi2015d,Gouveia2016b,Cambo2017,Zhao2017b,Kooiman2018b,Simons2018d,Mendoza2017,Morrison2017,Hochsmann2019e,Tong2019b,Klasnja2019,VanBlarigan2019b,Wernbacher2020,Kim2018c}, and Contribution metrics (e.g., number of User-Generated Content (UGC) posts) \cite{Hebden2014a,Liu2016,Zhao2017b,Althoff2017b,Mendoza2017,Yoon2018b,Korinek2018}. Inter-session metrics measure the user's loyalty to ST technology. They include Session to Session metrics, which capture the time between two sessions, and Session to Extended Period metrics, which capture longer periods, such as weeks, months, or years. Session to Session metrics \cite{Gouveia2014,Gouveia2015a,Gouveia2016b} include metrics, such as time between sessions (absence time) or MVPA between sessions. Session to Extended Period metrics \cite{Verbeek2012a,Hebden2014a,Bickmore2013,Gouveia2014,Gouveia2015a,AlAyubi2014,Zuckerman2014a,Washington2014a,Arigo2015,Garde2015,Blackman2015b,Katule2016b,Poirier2016a,Gouveia2016b,Monroe2017b,Leinonen2017b,Althoff2017b,Chung2017a,Kooiman2018b,Simons2018d,Shameli2019,Esakia2018c,Zhang2019a,Zhou2018c,Lyons2017b,Morrison2017,VanWoudenberg2020b,Babar2018b,Yoon2018b,Hochsmann2019e,Tong2019b,Nuijten2019,VanBlarigan2019b,Wernbacher2020,Korinek2018,Bianchi-Hayes2018b,Gremaud2018b,Polgreen2018b,Corepal2019b,Arrogi2019,Gell2020a,Cauchard2019b,Liew2020b,Middelweerd2020b}, include, among others, number of valid wear days, wear time and total usage. Metrics marked with an asterisk indicate that they are primarily targeted to wearable technology. Notice that the purpose of the SELF component and the respective documentation is not to provide definitions for all used metrics. Our goal is to provide practitioners and researchers with a summarized and concise idea of the metrics utilized in the investigated papers, most of which are widely used and are easily identified through the citation list or a search engine.
\begin{table}[htb!]
    \resizebox{.95\textwidth}{!}{
        \begin{tabular}{|a|aa}
        \hline
        \rowcolor[HTML]{3c3e69} 
        \multicolumn{3}{|c|}{\color[HTML]{FFFFFF} \textbf{BEHAVIORAL SELF ASPECT OF UE}}\\
        
            \hline
            \rowcolor[HTML]{787CD2} 
            \multicolumn{2}{|c|}{\cellcolor[HTML]{787CD2}{\color[HTML]{FFFFFF} \textbf{INTRA-SESSION (WITHIN $\rightarrow$ ACTIVITY)}}} & \multicolumn{1}{c|}{\cellcolor[HTML]{787CD2}{\color[HTML]{FFFFFF} \textbf{INTER-SESSION (ACROSS $\rightarrow$ LOYALTY)}}} \\ 
            \hline
            
            \rowcolor[HTML]{A1A3E0}
            \color[HTML]{FFFFFF}\textbf{Involvement} & \multicolumn{1}{l|}{\color[HTML]{FFFFFF}\textbf{Granularity}} & \multicolumn{1}{l|}{\color[HTML]{FFFFFF}\textbf{Session to Session}} \\

            \begin{tabular}[c]{@{}l@{}}\\\tabitem Dwell Time \\ \tabitem Session Duration\\ \tabitem Screen \& Page Views (Screen Tap \& Button Press Depth)\\ \tabitem Revisit Rate\\ \tabitem Glances Rate (Bounce Rate)\\\hfill\end{tabular} & 
            
            \multicolumn{1}{d|}{} &
            
            \multicolumn{1}{a|}{\begin{tabular}[c]{@{}l@{}}\tabitem Time between Sessions (Absence Time)\\ \tabitem Distanced Walked between Sessions\\ \tabitem Step Count between Sessions \\ \tabitem MVPA between Sessions\end{tabular}} \\ 
            \cline{1-1} \cline{3-3}
            
            \cellcolor[HTML]{A1A3E0}\color[HTML]{FFFFFF}\textbf{Interaction} & \multicolumn{1}{d|}{} & \multicolumn{1}{l|}{\cellcolor[HTML]{A1A3E0}\color[HTML]{FFFFFF}\textbf{Session to Extended Period}} \\ 
            \cline{1-1} \cline{3-3} 
            
            \begin{tabular}[c]{@{}l@{}}\\\tabitem Number of Content Shares, Comments \& Likes\\ \tabitem Number of Feature Accesses\\ \tabitem Number of Accepted Notifications (e.g., reminders, recommendations)\\ \tabitem Number of Content Views (e.g., messages, posts, e-mails)\\ \tabitem Notifications \& Content Response Time\\ \tabitem Time till PA\\ \tabitem Number of Watch App Accesses (*)\\ \tabitem Number of Watch Screen Swipes (*)\\ \tabitem Tapped Watch Screen Locations (*)\\\hfill\end{tabular} & 
            \multicolumn{1}{d|}{\begin{tabular}[c]{@{}c@{}}Feature\\ $\downarrow$\\ Page\\ $\downarrow$\\ Visit\\ $\downarrow$\\ Session\end{tabular}} & 
            
            \multicolumn{1}{a|}{\multirow[t]{3}{*}{\begin{tabular}[c]{@{}l@{}}\tabitem Number of Active \& Valid Wear Days\\ \tabitem Number of Sessions\\ \tabitem Total Usage \& Wear Time\\ \tabitem Number of Screen Taps \& Button Presses \\ \tabitem Number of Content Shares, Comments \& Likes\\\tabitem Number of Content Views \\ \tabitem Attrition Rate\\ \tabitem Number of PA Joined Challenges (IA)\\ \tabitem Number of Swapped Commodities \& Rewards (IA)\end{tabular}}}
            
            \\ 
            \cline{1-1}
            
            \cellcolor[HTML]{A1A3E0}\color[HTML]{FFFFFF}\textbf{Contribution} & \multicolumn{1}{a|}{} & \multicolumn{1}{a|}{} \\ 
            \cline{1-1}
            
            \begin{tabular}[c]{@{}l@{}}\\\tabitem Number of Set PA Goals\\ \tabitem Number of Replies (e.g., messages, EMA)\\ \tabitem Number of UGC Posts\\\hfill \end{tabular} & \multicolumn{1}{a|}{} & \multicolumn{1}{a|}{} \\ 
            \hline
            
        \end{tabular}
    }
    \caption{The \rev{Behavioral Self} Aspect of UE: Intra-session and inter-session metrics for quantifying user behavior in ST technology.\label{tab:behavioral}}
    \vspace{-6mm}
\end{table}

\subsubsection{The Environmental Aspect of UE}\hfill\\
Finally, SELF Table~\ref{tab:environmental} presents the \textbf{Environmental Aspect of UE}. It outlines the external constraints that the investigated papers have taken into account in the study of ST technology for HBC. In terms of \textbf{User Constraints}, articles have utilized the users' daily schedule \cite{Gouveia2014,Vathsangam2014,Rabbi2015e,Rabbi2015d,Cambo2017,Simons2018d,Ciravegna2019b}, daily commute patterns \cite{Gouveia2014,Ciravegna2019b}, anthropometrics \cite{Arigo2015,Rabbi2015e,Finkelstein2015e,Finkelstein2015d,Bronikowski2016,Klausen2016c,Chen2017,Monroe2017b,Lyons2017b,Leinonen2017b,Pope2018b,Zhou2018c,Gremaud2018b,Kooiman2018b,Polgreen2018b,Memon2018b,Pope2019b,Cadmus-Bertram2019b,Choi2019b,Hochsmann2019e,Tong2019b,Galy2019b,Zhang2019a,Cauchard2019b,VanWoudenberg2020b}, and cardiorespiratory fitness \cite{Klausen2016c,Chen2017,Pope2018b}. Regarding \textit{Social Constraints}, articles have used the users' physical and virtual social network \cite{Foster2010a,Khalil2013,Walsh2014a,Kramer2019,Chen2014b,Chen2016,AlAyubi2014,Liu2016,Leinonen2017b,Babar2018b}, as well as their social interactions \cite{Foster2010a,Walsh2014a,Chen2014b,Chen2016,AlAyubi2014,Arigo2015,Mendoza2017,Althoff2017b,Chung2017a,Babar2018b}. Also, \textit{Geographical Constraints} have been utilized to contextualize user content and explain user behavior, including geolocation \cite{Gouveia2014,Rabbi2015e,Rabbi2015d,Gouveia2015a,Blackman2015b,Poirier2016a,Leinonen2017b,Morrison2017,Simons2018d,VanDantzig2018c,Elliott2019b,Ciravegna2019b,Klasnja2019}, Indoor Location \cite{Cambo2017,VanDantzig2018c}, visited places \cite{Gouveia2014,Poirier2016a}, and weather \cite{Washington2014a,VanDantzig2018c,Klasnja2019}. Finally, in terms of \textit{Time Constraints}, time and day of the week \cite{Klasnja2019}, as well as bank holidays \cite{Klasnja2019} have been utilized in the investigated articles. Various studies in our article pool reported differences in performed PA between interventions conducted during spring versus interventions conducting in winter, as well as drops in PA during bank holidays. Such findings highlight the effect a user's environment has on HBC and performed PA; hence, it should be taken into consideration for future research.
\begin{table}[htb!]
\resizebox{.45\textwidth}{!}{%
\begin{tabular}{|Sl|Sl|}
\hline
\rowcolor[HTML]{3C3E69} 
\multicolumn{2}{|c|}{\cellcolor[HTML]{3C3E69}{\color[HTML]{FFFFFF} \textbf{ENVIRONMENTAL ASPECT OF UE}}} \\ \hline
\rowcolor[HTML]{A1A3E0} 
{\color[HTML]{FFFFFF} \textbf{User Constraints}} & {\color[HTML]{FFFFFF} \textbf{Geographical Constraints}} \\ 
\rowcolor[HTML]{F2F2FB} 

\begin{tabular}[c]{@{}l@{}}\tabitem Daily Schedule\\ \tabitem Daily Commute\\ \tabitem Anthropometrics\\ \tabitem Cardiorespiratory Fitness\end{tabular} & 

\begin{tabular}[c]{@{}l@{}}\tabitem Geolocation\\ \tabitem Indoor Location\\ \tabitem Visited Places\\ \tabitem Weather\end{tabular} \\ \hline

\rowcolor[HTML]{A1A3E0} 
{\color[HTML]{FFFFFF} \textbf{Social Constraints}} & {\color[HTML]{FFFFFF} \textbf{Time Constraints}} \\ 
\rowcolor[HTML]{F2F2FB} 

\begin{tabular}[c]{@{}l@{}}\tabitem Social Network\\ \tabitem Social Interactions\end{tabular} &

\begin{tabular}[c]{@{}l@{}}\tabitem Time and Day of the Week\\ \tabitem Holidays\end{tabular} \\ \hline

\end{tabular}%
}
\caption{The Environmental Aspect of UE: User, social, geographical and time constraints related to UE.\label{tab:environmental}}
\vspace{-6mm}
\end{table}

\subsection{A Use Case Scenario for the PAST SELF Framework\label{usecases}}
Now that we have showcased both components of the PAST SELF Framework in Sections~\ref{periodicTable} and \ref{self}, we proceed to demonstrate how practitioners and researchers can apply them in practice. To this end, we present in Figure~\ref{fig:example} a use case scenario with the timeline of a regular workday of an ST technology user. The timeline consists of five distinct lanes. The first four represent the four aspects of UE, and the last one includes the design techniques utilized in the development of an ST technological product. The dotted lines represent a timestamped event in our scenario. The parallel timelines showcase how the PAST component interacts with the SELF component and how UE's different aspects are interwoven. As stated previously, our scenario assumes a day in an ST user's life, which consists of various groups of events:

\begin{figure}[htb!]
  \centering
  \includegraphics[width=.95\textwidth]{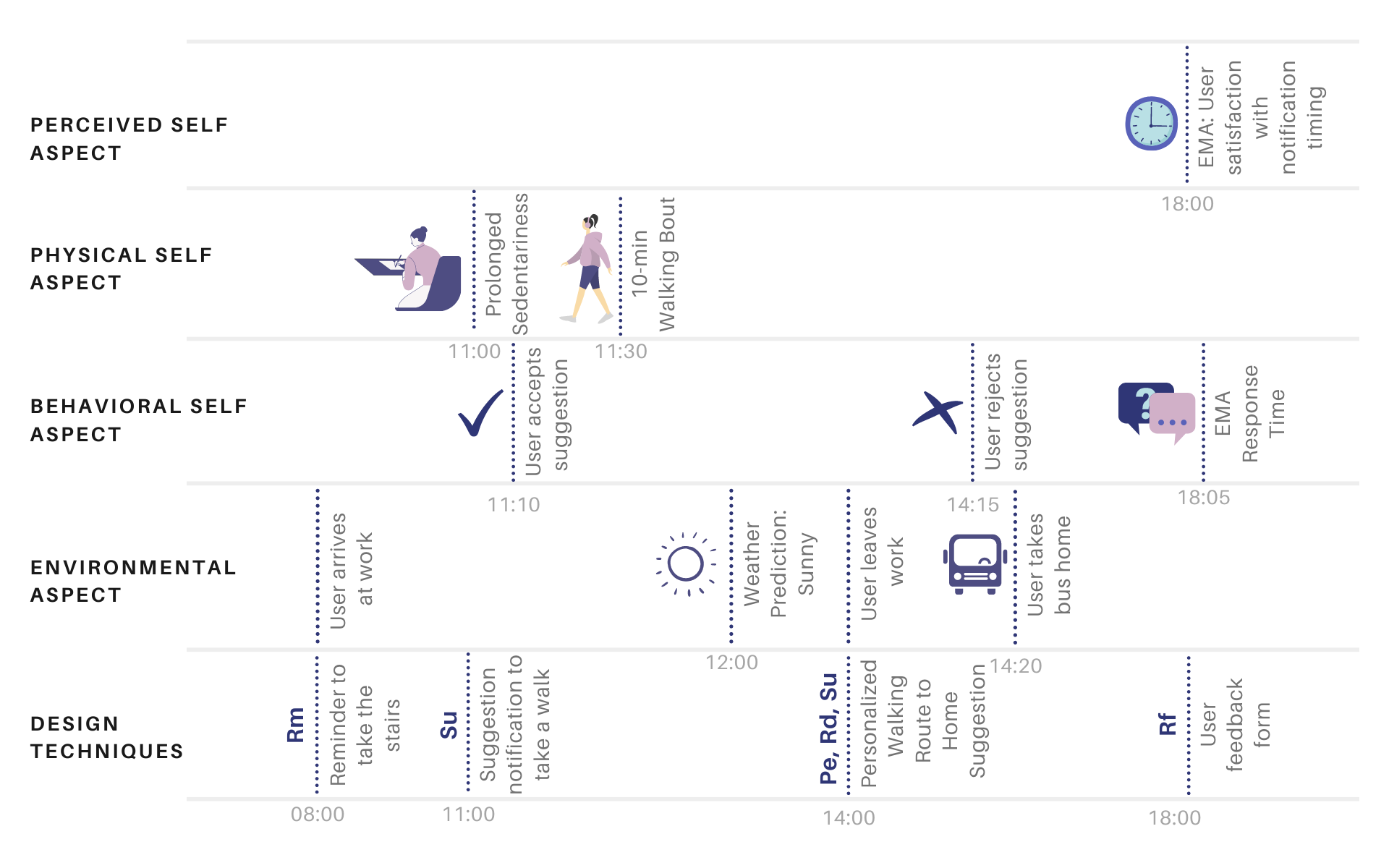}
  \caption{A use case scenario of how the PAST SELF framework can be applied in practice.}
  \Description{An example use case scenario of our PAST SELF framework. The parallel timelines showcase how the PAST component interacts with the SELF component and how UE's different aspects are interwoven.}
  \label{fig:example}
\end{figure}

\begin{itemize}[leftmargin=*]
    \item At 08:00, the user arrives at work. The system identifies the change in the user's location (Environmental Aspect of UE). It then sends the user a reminder on their wearable device to take the stairs instead of the elevator (Reminder Design Technique).
    \item After a few hours, at 11:00, the system detects prolonged sedentary activity (\rev{Physical Self} Aspect of UE). Thus, a push notification from their ST app suggests that the user take a short walk (Suggestion Design Technique). Upon seeing the notification, the user accepts the prompt (\rev{Behavioral Self} Aspect of UE) and completes a 10-min walking bout by 11:30 (\rev{Physical Self} Aspect of UE). The user's acceptance can help the system improve itself by learning the appropriate interruption times based on the user's schedule.
    \item Based on a sunny weather prediction for the afternoon, the system recommends a personalized route to home for the user (Personalization, Reduction \& Suggestion Design Techniques). The recommendation happens upon identifying a change in the user's environment at 14:00 when the user leaves work. However, the user soon rejects the prompt (Behavioral Self Aspect of UE) and takes a bus (Environmental Aspect of UE). The ST technology developer can interpret this rejection in various ways; the user might go to the gym after work rather than home, or the suggested route might be too long or too dangerous for the user. In any case, identifying this user behavior can help the system improve itself by suggesting alternative routes or sending earlier notifications to help the user mentally prepare for this schedule change.  
    \item Finally, at 18:00, the system sends a Momentary Ecological Assessment (EMA) to the user, which takes the form of a push notification from the ST app. The EMA's goal is to evaluate the user's satisfaction levels with today's notifications' timing. The user completes the short survey prompt by 18:05, which means a 5-min EMA response time (Behavioral Self Aspect of UE). This fast response time might signify that early evening might be convenient for the user to complete short feedback forms (Real-world Feel Design Technique).
\end{itemize}

\section{Discussion \& Conclusions\label{conclusions}}
This paper provides a detailed systematic review of 129 articles through which we identified the requirements of ST technology, namely successful HBC and UE, and the limitations and challenges of existing systems, such as dubious effectiveness and high attrition rates. Most importantly, we identified a lack of a formal framework that showcases the critical dimensions in the design and evaluation of ST technology. The absence of a common framework makes it harder to compare different studies in our article pool due to the various methodologies used for their design and evaluation. Hence, we organize the literature review around HBC and UE's key dimensions and propose a generic, comprehensive framework, PAST SELF, for the design and evaluation of ST technology. \rev{We believe that this review of articles and their analysis, the PAST SELF framework, and the open data and tool can be helpful for researchers and practitioners of ubiquitous ST technology. Below, we summarize our findings and discuss insights, limitations, and open challenges for future research:}

\noindent\rev{\textbf{HCI Design \& PSD Component Effectiveness:}} \rev{As illustrated in our findings, all PSD design techniques in ST systems can be effective under different settings, be that cohort characteristics, intervention duration, or sample size. At the same time, no single technique can guarantee success individually. It is the harmonic and tailored combination of different features that can encourage HBC for individuals. Nevertheless, some techniques have been used in practice more frequently than others. While commonly used techniques, such as Goal Setting or Self-monitoring, are proven to have high effectiveness, infrequent but promising techniques, such as Similarity and Variability, deserve more scientific attention. We believe that this review and its comprehensive presentation of HCI feature design can guide ST researchers and practitioners in choosing the PSD techniques that are more suitable for their use case and implementing them in real-world systems.}
    
\noindent\rev{\textbf{HCI Evaluation \& UE Quantification:}} \rev{While UE with a single system does not necessarily imply sustained HBC, it is evident that unsuccessful or ultimately abandoned ST technology cannot assist the user in their HBC journey. To monitor and improve the quality of the HCI with the ST system, we need a way to quantify UE. However, UE is a multi-faceted concept that goes beyond simply measuring PA or health changes. It encompasses various aspects of the user's persona and interaction with the system, such as the user’s thoughts and affective states, health habits and behavioral patterns, or the environmental factors that affect the user’s interaction with the system. This review provides a comprehensive list of multi-faceted UE evaluation metrics that have been used in the related literature. Its goal is to assist ST researchers and practitioners with accompanying the users on their ST journey through constant, iterative evaluation and adaptation of the ST system.}

\noindent\rev{\textbf{Sample Size \& Intervention Duration:}} The majority of the investigated studies have limited sample size and duration, which might undermine the generalizability of their results and highlights a need for large-scale studies. \rev{ However, it is important to note that the majority of the studies are restricted by the time and budget available to the researchers; hence small-scale studies are more realistic given such restrictions. Additionally, small-scale studies are more suitable for rapid prototyping and testing in the early stages of the development of ST technology solutions for HBC. It might take several years to demonstrate sustained HBC, but surely, quantifying indications of HBC in the short term can be more straightforward. Nevertheless, in later stages of technology development, large-scale studies can help build more robust and reliable ST technology. Moreover, they enable the collection of greater amounts of data for service tailoring and personalization through big data analytics and A/B testing.}
    
\noindent\rev{\textbf{Cohorts, Equity \& Accessibility:} As our results illustrate, there is a lack of research on how different subgroups of users (e.g., race, age, gender, health status) interact with ST technology, presenting an opportunity for future work. Future research in the field should focus on equitable access to ST technology for HBC, taking into account the socioeconomic status, variable health, and technological literacy of cohorts, while being sensitive to their cultural and language needs. Similarly, a promising direction for future research is the accessibility of ST technology, which we have not encountered in the included primary studies. Accessibility should go beyond addressing physical usability to an in-depth analysis of the population's needs in terms of disabilities. It is important to note that such user subgroups frequently come from disadvantaged or minority population segments; hence research in the field should always raise questions concerning ethical concepts.}
    
\noindent\textbf{The PAST SELF Framework:} Our article pool utilizes a variety of design and evaluation techniques and metrics. This lack of standardization makes it difficult to compare results, evaluate each work's contribution, and obtain insights. Our PAST SELF framework brings together techniques utilized in previous works and comprehensively organizes them to guide future studies. Specifically, we propose the \textbf{P}eriodic T\textbf{a}ble of \textbf{S}elf-\textbf{T}racking Design (PAST), which showcases the most common design techniques for ST technology along with their expected efficacy. Also, we propose the \textbf{S}elf-Tracking \textbf{E}va\textbf{l}uation \textbf{F}ramework (SELF) which presents a comprehensive list of evaluation metrics organized under four key dimensions of UE. \rev{PAST SELF can help ST practitioners and researchers design and quantify the user experience to make more informed decisions for future interventions. 
Nevertheless, the role of the PAST SELF framework is, in reality, complementary. We recommend that interested parties should combine knowledge from our review and framework, as well as domain experts and the users' themselves through participatory design and citizen science practices.}
    
\noindent\rev{\textbf{Open Science \& Open Data:} To ensure the PAST SELF framework's maintainability and abide by the FAIR Data Principles \cite{wilkinson2016fair}, we make our corpus of primary studies publicly available through GitHub \cite{sofia_yfantidou_2020_4063377}. Our aim for this repository is twofold: (i) It can serve as a live dataset of ST technology interventions for HBC and UE, where new studies will be added, thus keeping up with technological advances in ST for HBC. To this end, we encourage researchers and practitioners to share their work with the research team and contribute to this public repository. (ii) Due to its detailed information, our dataset can facilitate further experimentation, analysis, and derivation of additional metrics.}

\noindent\rev{\textbf{Online Exploration Tool.} To overcome the pitfalls of static reporting, we have created an online, interactive exploration tool for the PAST SELF framework\footnote{"PAST SELF Framework Tool." \url{https://syfantid.github.io/past-framework-visualization/}. Accessed 4 Jan. 2021.}. This tool enables researchers and practitioners to adjust the PAST component in real-time according to their application requirements through various filtering and weight adjustment options. This way, a researcher can tailor the PAST score according to their application needs and for specialized cohorts, or they can choose to favor large-scale, long-term studies over small-scale, short-term studies and the opposite. Such filtering options eliminate the potential danger of authors' critique or biases affecting the reported PAST\_score. On the contrary, through the interaction with the tool, the user can explore if and how different weights and filters affect the positioning of persuasive techniques in the PAST table.} 
    
Surely, the entire field of ST for HBC is quite broad and cannot be covered by a single literature review. While our review primarily focuses on the design and evaluation aspects for the general population, future studies could focus on exploring the literature related to different sample groups to provide specialized versions of the PAST SELF framework for each user segment with varying characteristics. \rev{Our adaptive exploration tool is the first step towards this direction.} Future studies could also concentrate on HBC's different aspects, such as stress management, smoking cessation, or disease control. PA is only the beginning of the possibilities ST devices have in capturing human data. Moreover, due to the lack of evaluation standardization, the strict criteria required to conduct a meta-analysis would significantly limit our article pool. \rev{For example, various studies did not report baselines before the intervention, while others did not report results for both control and intervention groups.} However, given a standardized reporting format in the future, a meta-analysis could provide valuable information about the efficacy of different ST design techniques. 
Finally, this review has left out some cutting-edge aspects of ST technology that are still not present in the related literature. For instance, only a limited number of papers utilize ML models for personalizing the user experience is ST. Hence, a categorization of such methods would not make sense at this point. However, future studies should focus on taking advantage of the state-of-the-art in ML to provide more customized ST products to the users. At the same time, new types of sensors and related functionality are incorporated into ST devices, such as ECG, fall detection, SpO$_2$ sensors, and integration with popular voice assistants, which may transform the field of ST into a complete health tracking experience. Future work should study if these advancements affect the users' HBC journey and UE with the technology itself.


%
\begin{acks}
This project has received funding from the European Union’s Horizon 2020 research and innovation programme under the Marie Skłodowska-Curie grant agreement No 813162. The content of this paper reflects only the authors' view and the Agency and the Commission are not responsible for any use that may be made of the information it contains. The authors would also like to thank T. Valk and S. Karamanidis, for their contribution to the development of the PAST SELF tool.
\end{acks}

\bibliographystyle{ACM-Reference-Format}
\bibliography{bibliography}

\end{document}